\documentclass[journal]{IEEEtran}

\usepackage{graphicx}
\usepackage{multirow}
\usepackage{textcomp}
\usepackage{array}
\usepackage[caption=false,font=footnotesize]{subfig}

\usepackage{algorithm}
\usepackage{algorithmic}

\usepackage{amsthm}
\theoremstyle{definition}

\usepackage{amsmath,amssymb,amsfonts}
\DeclareMathOperator*{\argmax}{argmax} 

\usepackage[cmintegrals]{newtxmath}
\newcolumntype{M}[1]{>{\centering\arraybackslash}m{#1}}

\usepackage{cite}

\begin{document}

% --- 縮減數學公式前後的空白 ---
\setlength{\abovedisplayskip}{3pt}
\setlength{\belowdisplayskip}{3pt}
\setlength{\abovedisplayshortskip}{3pt}
\setlength{\belowdisplayshortskip}{3pt}

% --- 縮減圖片/表格與內文之間的空白 ---
\setlength{\textfloatsep}{6pt plus 2pt minus 2pt}
\setlength{\floatsep}{6pt plus 2pt minus 2pt}
\setlength{\intextsep}{6pt plus 2pt minus 2pt}

\title{Curriculum-Guided Reinforcement Learning for Energy-Efficient UAV-ISAC in Post-Disaster Search-and-Rescue Operations}

\author{Tai-You~Guo and Chuan-Chi~Lai,~\IEEEmembership{Member,~IEEE}%
\thanks{This work was supported by the National Science and Technology Council (NSTC), Taiwan, under Grant No. NSTC 115-2221-E-194-042-MY2, and was also supported in part by the Advanced Institute of Manufacturing with High-tech Innovations (AIM-HI) from the Featured Areas Research Center Program within the framework of the Higher Education Sprout Project by the Ministry of Education (MOE) in Taiwan. \emph{(Corresponding author: Chuan-Chi~Lai.)}}%
\thanks{Tai-You Guo is with the Department of Communications Engineering, National Chung Cheng University, Minxiong Township, Chiayi County 621301, Taiwan.}%
\thanks{Chuan-Chi Lai is with the Department of Communications Engineering, National Chung Cheng University, Minxiong Township, Chiayi County 621301, Taiwan, and also with the Advanced Institute of Manufacturing with High-tech Innovations (AIM-HI), National Chung Cheng University, Minxiong Township, Chiayi County 621301, Taiwan.}%
}

\markboth{Preprint for IEEE Journal}%
{}

\IEEEtitleabstractindextext{%
\begin{abstract}
Uncrewed aerial vehicles (UAVs) are promising platforms for integrated sensing and communication (ISAC), but their limited onboard energy creates a strong coupling among sensing accuracy, communication quality, and propulsion cost. This paper proposes a curriculum-guided soft actor-critic (CG-SAC) framework with propulsion-aware reward shaping for energy-efficient UAV-ISAC, jointly optimizing the 3D trajectory, communication-sensing power split, and per-user power allocation. A rotary-wing propulsion model is incorporated to derive a closed-form propulsion-economic cruising speed, which is used to construct a propulsion-aware speed-shaping term within a normalized composite reward together with navigation, node-visiting, energy-efficiency, and constraint-penalty terms. A log-linear curriculum progressively tightens the communication, sensing, and proximity requirements during training. Across 2000 randomized scenarios, CG-SAC achieves an average energy efficiency of 0.72 Mbits/J, substantially outperforming the evaluated DRL baselines. Among successfully completed missions, it requires 107.6 steps on average, corresponding to a 66\%--82\% reduction in flight steps relative to the baselines. Crucially, the learned policy exhibits mission-aware speed adaptation by decelerating near service points and accelerating during transit, while achieving a 99.6\% communication-rate satisfaction ratio at service instants. Ablation results further demonstrate the complementary roles of the reward components in balancing mission feasibility and energy efficiency.
\end{abstract}

\begin{IEEEkeywords}
Uncrewed aerial vehicle (UAV), integrated sensing and communication (ISAC), energy efficiency, trajectory optimization, deep reinforcement learning, soft actor-critic, curriculum learning.
\end{IEEEkeywords}}

\maketitle
\IEEEdisplaynontitleabstractindextext
\IEEEpeerreviewmaketitle

\section{Introduction}
\label{sec:introduction}

\IEEEPARstart{A}{s} sixth-generation (6G) wireless systems evolve toward integrated sensing and communication (ISAC), the joint provision of communication and sensing services has emerged as an important direction for intelligent wireless networks \cite{lu2024tenchallenges,zhang2025intelligentisac}. By sharing spectrum and radio-frequency hardware, ISAC enables a common platform to support both connectivity and environmental awareness. This capability is particularly attractive for uncrewed aerial vehicles (UAVs), whose mobility and favorable line-of-sight (LoS) propagation can provide rapidly deployable sensing and communication services in areas where terrestrial infrastructure is unavailable or damaged \cite{meng2024uavisac,hu2025uavisacsurvey}. Among such applications, post-disaster search-and-rescue is especially compelling: a UAV can search for potential victims while simultaneously delivering the acquired information to ground rescue teams.

\begin{figure}[!t]
    \centering
    \includegraphics[width=.5\textwidth]{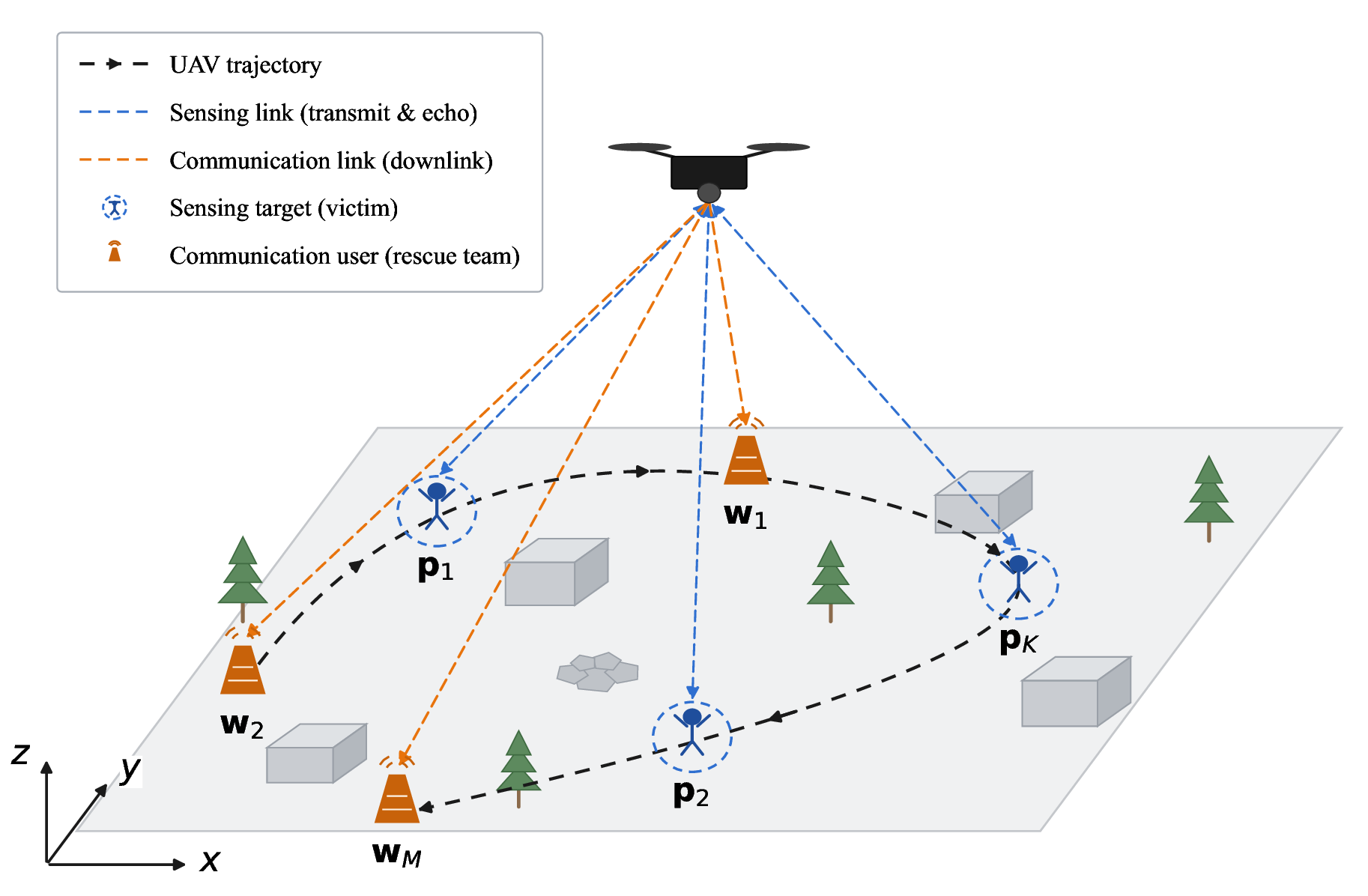}
    \caption{UAV-ISAC mission for post-disaster search-and-rescue operations.}
    \label{fig:rescue_scenario}
\end{figure}

In a post-disaster environment, however, the UAV is not merely required to establish wireless links or perform sensing at isolated locations. It must complete a mission over a sequence of spatially dispersed targets and communication users under a limited onboard energy budget. As illustrated in Fig.~\ref{fig:rescue_scenario}, the UAV jointly performs sensing of rescue targets $\mathbf{p}_1,\dots,\mathbf{p}_K$ and downlink communication to ground rescue teams $\mathbf{w}_1,\dots,\mathbf{w}_M$ using a common transmit waveform. The resulting mission requires the UAV to determine where to fly, how fast to fly, and how to allocate its transmit power between communication and sensing while visiting the required nodes. Hence, the objective is inherently mission-oriented rather than being limited to optimizing an individual communication or sensing metric.

The main difficulty arises from the strong coupling among mobility, sensing, communication, and propulsion energy. First, UAV propulsion accounts for a substantial portion of the onboard energy consumption, making inefficient flight trajectories or unnecessarily high speeds costly. Moreover, the propulsion cost depends nonlinearly on the flight speed, so minimizing mission duration does not necessarily minimize energy consumption. Second, sensing and communication share a limited transmit-power budget: allocating more power to sensing improves the sensing quality but leaves less power for downlink communication. Third, improving communication or sensing quality generally requires favorable three-dimensional positioning, which may increase the flight distance or force the UAV to deviate from an energy-efficient trajectory. These interactions create a tightly coupled continuous-control problem in which mission completion, service quality, and propulsion efficiency must be addressed simultaneously.

This coupling also makes direct reinforcement-learning optimization challenging. The action space contains continuous mobility and radio-resource decisions, while the mission-completion objective is sparse and the communication and sensing requirements impose stringent feasibility boundaries. A policy that focuses only on rapid node visiting may adopt unnecessarily aggressive flight speeds, whereas a policy that prioritizes service quality may spend excessive energy on prolonged or inefficient maneuvers. Therefore, an effective learning framework should not only explore the continuous control space, but also exploit the physical structure of UAV propulsion and progressively acquire feasible behaviors under increasingly stringent service requirements.

To address these issues, we propose a \textit{curriculum-guided soft actor-critic} (CG-SAC) framework with \textit{propulsion-aware reward shaping} for energy-efficient UAV-ISAC. The proposed framework jointly optimizes the 3D UAV trajectory, the communication-sensing power split, and the per-user power allocation. A speed-dependent rotary-wing propulsion model is first used to derive a closed-form economic cruising-speed reference. Rather than treating this quantity as a standalone optimization target, we incorporate the resulting propulsion insight into a speed-shaping term within a normalized composite reward. In parallel, a log-linear curriculum progressively tightens the communication, sensing, and proximity requirements during training, allowing the agent to first learn basic spatial navigation and subsequently adapt to stricter mission constraints. The resulting design couples physical energy awareness with curriculum-based exploration while retaining the maximum-entropy characteristics of SAC.

Under the proposed framework, the UAV sequentially visits the sensing targets and communication users, dynamically adapting its 3D motion and radio resources according to the spatial configuration and service requirements. In this way, the proposed method aims to learn a mission-aware policy that accelerates during transit while slowing down near service points when additional interaction time is beneficial. The sensing requirement is represented by an SNR-based sensing-error surrogate derived from the adopted echo model, while communication quality is characterized by the achievable downlink rate.

The main contributions of this paper are summarized as follows.
\begin{itemize}
	\item \textit{Mission-oriented joint optimization for energy-efficient UAV-ISAC.}
	We formulate a coupled optimization framework that jointly considers 3D trajectory control, communication-sensing power splitting, and per-user power allocation under propulsion-energy, communication-QoS, sensing-accuracy, and kinematic requirements. The formulation explicitly captures the interaction between spatial mobility, radio-resource allocation, and mission-level energy efficiency.

	\item \textit{Propulsion-aware reward shaping for continuous UAV control.}
	A rotary-wing propulsion model is used to derive a closed-form horizontal propulsion-economic cruising-speed reference. This physical insight is incorporated into the composite reward through a speed-shaping term. The propulsion-aware speed term provides an additional physical prior for shaping the learned flight-speed behavior, encouraging energy-efficient flight while retaining the flexibility to slow down near service locations and accelerate during transit.

	\item \textit{Curriculum-guided SAC with synchronized entropy scheduling.}
	A log-linear curriculum progressively tightens the communication, sensing, and proximity requirements during training. The curriculum determines task difficulty, while the entropy lower bound regulates exploration during curriculum transitions. Combined with propulsion-aware reward shaping, the proposed CG-SAC framework mitigates exploration difficulty in the highly constrained continuous-control space.

	\item \textit{Comprehensive evaluation under randomized mission conditions.}
	Extensive simulations over $2000$ randomized scenarios evaluate mission completion, energy efficiency, service quality, reward-component ablations, and sensitivity to the onboard energy budget. The results show that CG-SAC substantially improves the evaluated energy-efficiency and mission-completion performance over representative continuous-control DRL baselines, while maintaining high communication-rate satisfaction at service instants.
\end{itemize}

The remainder of this paper is organized as follows. Section~\ref{sec:related_works} reviews the related work. Section~\ref{sec:system} establishes the UAV-ISAC system model, and Section~\ref{sec:problem} formulates the joint optimization problem. Section~\ref{sec:method} presents the proposed curriculum-guided SAC framework. Section~\ref{sec:simulation} reports the simulation results, and Section~\ref{sec:conclusion} concludes the paper.

\begin{table*}[!t]
\centering
\caption{Comparison between this work and representative studies}
\label{tab:related_work}
\footnotesize
\setlength{\tabcolsep}{4pt}
\renewcommand{\arraystretch}{1.25}
\begin{tabular}{@{}l p{3cm} p{2.8cm} p{1.8cm} p{4cm} p{3cm}@{}}
\hline
\textbf{Reference (Year)} & \textbf{System scenario} & \textbf{Objective} & \textbf{Sensing metric} & \textbf{Jointly designed variables} & \textbf{Methodology} \\
\hline
\cite{9054054} (2020) & UAV-aided WSN data collection (non-ISAC) & Propulsion energy minimization & N/A & 3D flight trajectory & Convex optimization \\
\cite{10275015} (2024) & Cellular-connected UAV (non-ISAC) & Energy efficiency & N/A & Trajectory with map reconstruction & Deep reinforcement learning \\
\cite{10419470} (2023) & UAV-aided ISAC network & Age of information (AoI) minimization & N/A & Trajectory and service order & Dynamic programming \\
\cite{10571306} (2024) & UAV-ISAC with user-position uncertainty & Sensing-communication trade-off & CRB & Trajectory geometry & Artificial potential field \\
\cite{11329133} (2026) & Multi-UAV multi-static UAV-ISAC & Joint resource and trajectory design & Sensing metric & Trajectory, communication/sensing power split & Hierarchical multi-agent DRL \\
\hline
\textbf{This work} & \textbf{3D single-UAV ISAC (disaster relief)} & \textbf{EE maximization under QoS and sensing constraints} & \textbf{CRB-inspired sensing-error surrogate} & \textbf{3D trajectory, communication/sensing power split, per-user power allocation} & \textbf{Curriculum-guided SAC} \\
\hline
\end{tabular}
\vspace{-1em}
\end{table*}

\section{Related Work}
\label{sec:related_works}

ISAC aims to achieve data transmission and environmental sensing cooperatively under shared spectrum and hardware. Early studies focused mainly on the interference and performance trade-offs induced by physical-layer waveform design. The work in~\cite{11074935} investigated \textit{frequency-hopping multiple-input multiple-output} (FH-MIMO) architectures in \textit{dual-function radar-communication} (DFRC) systems and showed that embedding communication symbols through \textit{frequency-hopping code selection} (FHCS) introduces undesirable sidelobes in the radar ambiguity function, thereby degrading the sensing resolution. To quantify this physical limit, the authors of~\cite{10251151} derived the Pareto-optimal boundary between the Cram\'{e}r-Rao Bound (CRB) and the achievable rate in multi-antenna ISAC systems, confirming that high-accuracy sensing is necessarily accompanied by a loss of communication throughput. As system complexity grows, auxiliary techniques and multi-dimensional resource allocation have also received attention. The study in~\cite{10364735} addressed the joint beamforming and reflection-matrix design for \textit{reconfigurable intelligent surface} (RIS)-aided ISAC, whereas~\cite{11165372} proposed a predictive \textit{integrated sensing, communication, and computation over-the-air} (ISCCO) framework for vehicular networks, in which an \textit{extended Kalman filter} (EKF) assists downlink beamforming and the convergence stability of reinforcement learning in dynamic resource allocation is preliminarily verified.

Given the limited onboard battery capacity of a UAV, accurate propulsion power modeling serves as the physical foundation for trajectory optimization. A closed-form energy model covering initial speed, acceleration, and flight time for rotary-wing UAVs was proposed in~\cite{9461176}, improving upon earlier constant-speed assumptions. Through flight experiments, the authors of~\cite{9495369} decomposed the energy consumption into induced, profile, and parasitic power, and revealed a pronounced U-shaped curve relating EE to flight speed. The motion characteristics of a UAV also directly shape its energy profile: based on multi-rotor dynamics,~\cite{9902793} showed that frequent vertical maneuvers and horizontal accelerations substantially raise the energy cost, while~\cite{10403152} evaluated a \textit{proton-exchange-membrane fuel-cell} (PEMFC) propulsion system and reduced the parasitic consumption through an integrated cathode-flow-channel design with airflow cooling. Collectively, these results indicate that dynamic propulsion energy must be treated as a primary consideration in UAV-ISAC trajectory planning.

For highly nonlinear and time-varying environments, DRL offers advantages in handling high-dimensional non-convex optimization. In UAV-assisted \textit{mobile edge computing} (MEC), \textit{model-agnostic meta-learning} (MAML) was combined with DRL in~\cite{11121871} to accelerate environmental adaptation; Lyapunov optimization was integrated with DRL in~\cite{10896833} to guarantee long-term energy constraints under stochastic task arrivals; and \textit{multi-agent proximal policy optimization} (MAPPO) was applied in~\cite{10412167} to jointly configure 3D trajectories, \textit{non-orthogonal multiple access} (NOMA) power, and spectrum resources. Regarding stability and safety, existing studies continue to refine the actor-critic architecture originating from DDPG~\cite{lillicrap2016ddpg}. TD3~\cite{fujimoto2018td3} may exhibit insufficient exploration or degraded stability under tightly constrained continuous control, which was mitigated in~\cite{10704600} to optimize the EE of RIS-aided networks, and Safe-TD3 was proposed in~\cite{10330124} to convert energy limits into hard safety constraints. A broader account of how these algorithm families have been applied to UAV missions can be found in the recent survey~\cite{amodu2025drlsurvey}. However, when the sensing accuracy requirement of an ISAC system is both tightly coupled and stochastic, the exploration capability of deterministic policies in a complex action space remains insufficient.

Table~\ref{tab:related_work} positions this work against representative studies. As shown, existing efforts rarely reconcile ISAC operation with EE in trajectory design. Specifically,~\cite{9054054} and~\cite{10275015} build realistic propulsion energy models and pursue energy-efficient trajectories through conventional optimization and DRL, respectively, yet neither considers an ISAC network nor the joint design of sensing and communication functionalities. Conversely, existing UAV-ISAC trajectory studies largely neglect energy utilization: the work in~\cite{10419470} minimizes the AoI and determines the service order by dynamic programming, and~\cite{10571306} combines an artificial potential field to cope with user-position uncertainty; neither treats EE as a core objective nor addresses the communication-sensing power split. However, \cite{11329133} focuses on multi-UAV cooperation and hybrid discrete/continuous control, whereas the present work considers single-UAV mission-level energy efficiency with continuous control.

In summary, although prior studies have separately advanced ISAC, energy modeling, trajectory design, and reinforcement learning, most remain confined to a single facet and have yet to integrate sensing, communication, EE, trajectory planning, and resource allocation into one tightly coupled non-convex problem. This is particularly true for 3D UAV-ISAC scenarios with variable altitude, in which the sensing accuracy (surrogate threshold), the ground QoS rate, and the propulsion energy (characterized by the U-shaped curve) compete and interact. To address this gap, this study adopts the maximum-entropy SAC algorithm and constructs an EE-centric joint optimization framework for UAV-ISAC that simultaneously considers the 3D trajectory, the communication-sensing power split, and per-user resource allocation, while incorporating a speed-dependent propulsion energy model and strict performance constraints.

\section{System Model}
\label{sec:system}

\subsection{Network Architecture}
We consider a UAV-ISAC system in 3D space, in which a rotary-wing UAV equipped with $N_t$ antennas acts as an aerial base station over a mission period of duration $T$. For tractability, $T$ is discretized into $N$ equal time slots of length $\delta t=T/N$, and the time-varying 3D coordinate of the UAV in slot $n\in\mathcal{N}=\{1,\dots,N\}$ is written as $\mathbf{q}[n]=[x[n],y[n],z[n]]^{T}\in\mathbb{R}^{3}$.

\begin{figure}[!t]
	\centering
	\includegraphics[width=.4\textwidth]{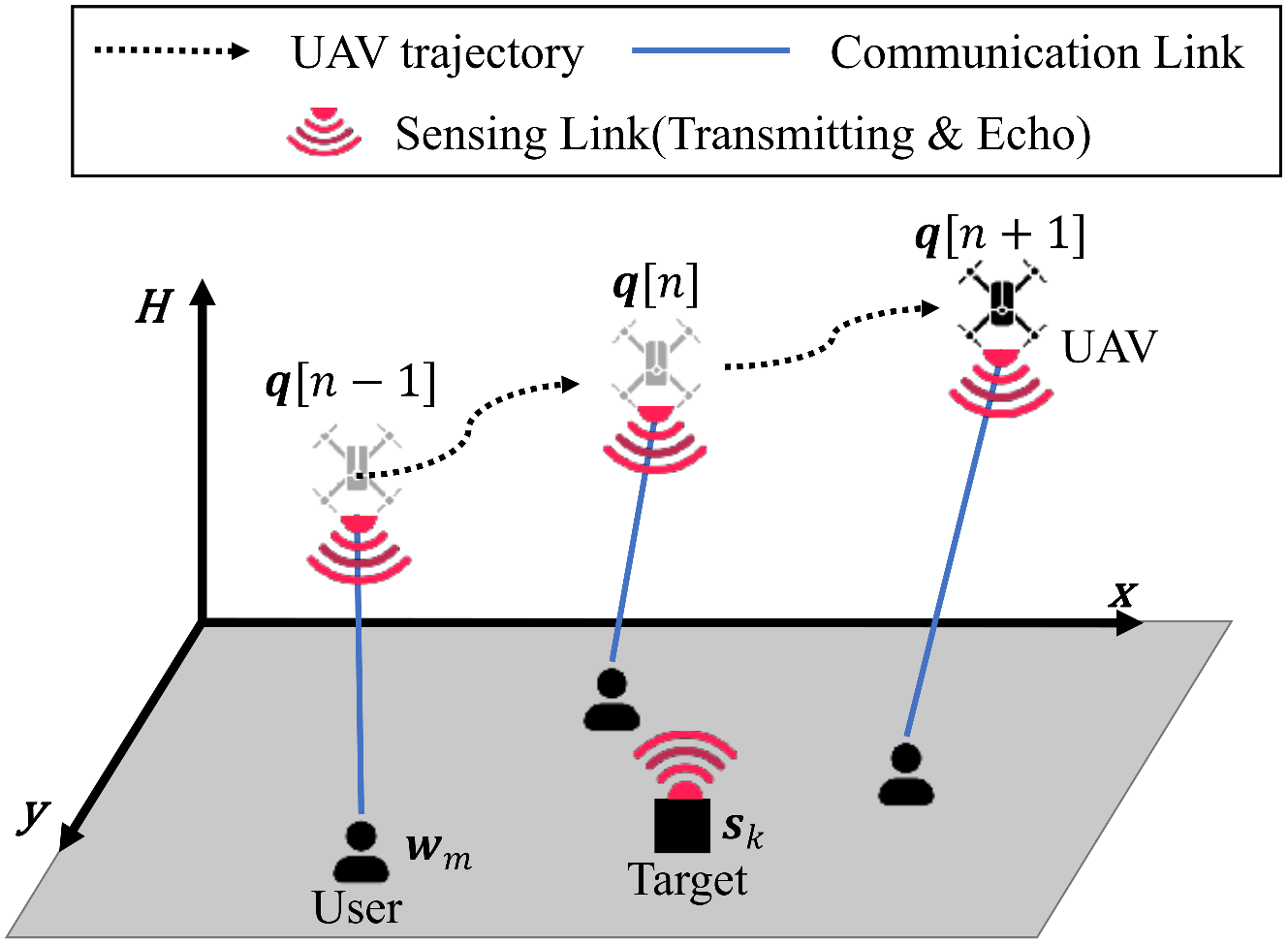}
	\caption{The considered UAV-ISAC system architecture.}
	\label{fig:system_model}
\end{figure}

The UAV employs a single transmit waveform to realize communication and sensing simultaneously. On the downlink, it serves $M$ single-antenna communication users (CUs) distributed on the ground, whose position vectors are $\mathbf{w}_m=[x_m,y_m,0]^{T}$, $m\in\{1,\dots,M\}$. In parallel, it exploits the echoes of its transmitted signal to estimate the parameters of $K$ sensing targets located at $\mathbf{p}_k=[x_k,y_k,0]^{T}$, $k\in\{1,\dots,K\}$. The overall architecture is depicted in Fig.~\ref{fig:system_model}.

To ground the model in a concrete application, we adopt the disaster-relief scenario introduced in Section~\ref{sec:introduction}: the sensing targets correspond to rescue points to be detected and localized (e.g., trapped survivors), and the communication users correspond to ground rescue teams awaiting notification. The UAV performs \emph{patrol-style sequential service}, flying to each node in turn to complete an accuracy-compliant localization at a sensing target or a rate-compliant information delivery at a communication user. Only when all rescue points have been localized \emph{and} all rescue teams have been notified is the mission considered complete; this is the physical meaning of adopting ``all nodes visited'' as the completion criterion. Accordingly, we set $M=K=3$ so as to model a representative small relief unit that balances scenario realism against training tractability, noting that the framework extends directly to a larger number of nodes. In practice, a large-scale disaster area can be geographically partitioned into multiple micro-grids or clusters, with a single UAV sequentially assigned to patrol each small-scale unit. This clustering strategy ensures that the action and state spaces remain tractable while preserving the physical realism of the aerodynamic and QoS constraints. 

Considering the limited onboard battery and computing resources, the UAV is assumed to be connected to a ground core base station equipped with an edge-computing server through a wireless backhaul link. In the DRL workflow, the computationally intensive training stage (experience replay and gradient updates) is offloaded to the ground server; the UAV only receives the updated actor-network weights and performs lightweight forward-pass inference onboard. Such an edge-assisted deployment effectively relieves the onboard computing unit.

\subsection{Channel Model}
Air-to-ground (A2G) links are adopted between the UAV and the ground users. Let $\mathbf{w}_m$ and $\mathbf{q}(t)$ denote the 3D positions of the $m$-th user and the UAV in slot $t$, so that their geometric distance is $d^{\mathrm{CU}}_m(t)=\|\mathbf{q}(t)-\mathbf{w}_m\|$. To focus on system-level trajectory and resource optimization, the random fluctuation of small-scale fading is neglected~\cite{10412167}, i.e., $\tilde{h}_m(t)=1$; the channel power gain $|h_m(t)|^2$ is thus fully determined by the reciprocal of the large-scale path loss $L_m(t)$, namely $|h_m(t)|^2=\rho_m(t)=1/L_m(t)$.

The occurrence of LoS and non-LoS (NLoS) propagation on an A2G link depends on the elevation angle $\theta^{\mathrm{CU}}_m(t)$ of the UAV as seen from the user. We adopt the widely used sigmoid LoS probability model originally proposed in~\cite{alhourani2014lap} and subsequently employed in UAV trajectory studies such as~\cite{10275015},
\begin{equation}
P_{\mathrm{LoS}}(\theta^{\mathrm{CU}}_m(t))=\frac{1}{1+\alpha_{\mathrm{L}}\exp\!\left(-\beta_{\mathrm{L}}\left[\theta^{\mathrm{CU}}_m(t)-\alpha_{\mathrm{L}}\right]\right)},
\end{equation}
where $\alpha_{\mathrm{L}}$ and $\beta_{\mathrm{L}}$ are environment-dependent parameters. The average path loss of the A2G channel is accordingly
\begin{equation}
L_m(t)=P_{\mathrm{LoS}}(\theta^{\mathrm{CU}}_m(t))L_{\mathrm{LoS}}(t)+\bigl(1-P_{\mathrm{LoS}}(\theta^{\mathrm{CU}}_m(t))\bigr)L_{\mathrm{NLoS}}(t),
\end{equation}
with $L_{\mathrm{LoS}}(t)$ and $L_{\mathrm{NLoS}}(t)$ denoting the path loss under LoS and NLoS conditions, respectively. This model captures how the 3D trajectory of the UAV dynamically shapes the link quality and serves as the basis of the subsequent resource allocation.

\subsection{Propulsion Energy Model and Economic Speed}
\label{subsec:energy}
The onboard energy limit is the central factor constraining the mission duration and the trajectory design. According to the closed-form rotary-wing propulsion model of~\cite{zeng2019rotarywing} and its subsequent refinements and experimental validation~\cite{9461176,9495369}, the propulsion power of a rotary-wing UAV is highly nonlinear and generally depends on its flight state and motion.

We therefore adopt a computationally light and smooth approximation that preserves the dominant speed dependence. Specifically, the propulsion power is modeled as a hovering baseline plus a parasitic term growing with the cube of the speed,
\begin{equation}
\label{eq:pfly}
\mathcal{P}_{\mathrm{fly}}(t)=P_{\mathrm{hover}}+\kappa\,\|\mathbf{v}_{xy}(t)\|^{3} + \kappa_z [v_z(t)]^+,
\end{equation}
where $\kappa=\frac{P_{\mathrm{full}}-P_{\mathrm{hover}}}{V_{\max}^{3}}$ is calibrated from the hovering power $P_{\mathrm{hover}}$, the full-speed power $P_{\mathrm{full}}$, and the maximum horizontal speed $V_{\max}$, and $\|\mathbf{v}_{xy}(t)\|$ is the horizontal speed; $\kappa_z$ is the proportionality coefficient for the vertical climbing power, and $[v_z(t)]^+ = \max(v_z(t), 0)$ ensures this additional energy cost is only counted when the UAV climbs. The total system power in slot $t$ is then
\begin{equation}
\mathcal{P}_{\mathrm{total}}(t)=\mathcal{P}_{\mathrm{fly}}(t)+P_{\max}+P_{\mathrm{elec}}.
\label{eq:ptotal}
\end{equation}

For the horizontal cruising-speed analysis, the vertical-climb term is omitted by assuming $v_z=0$. Although simplified, \eqref{eq:pfly} retains the essential trade-off between energy efficiency and flight speed. Measured in energy per unit horizontal displacement, it yields
\begin{equation}
\frac{\mathcal{P}_{\mathrm{fly}}(t)}{\|\mathbf{v}_{xy}(t)\|}=\frac{P_{\mathrm{hover}}}{\|\mathbf{v}_{xy}(t)\|}+\kappa\,\|\mathbf{v}_{xy}(t)\|^{2},
\end{equation}
which is a typical U-shaped curve: at very low speeds the temporal accumulation of the hovering power inflates the energy per distance, whereas at very high speeds the cubic growth of the parasitic power does the same. Minimizing this convex function yields the model-derived \emph{horizontal propulsion-economic cruising speed} $V_{\mathrm{eco}}$, which is used as a physically motivated speed reference for reward shaping,
\begin{equation}
V_{\mathrm{eco}}=\left(\frac{P_{\mathrm{hover}}}{2\kappa}\right)^{1/3}.
\label{eq:veco}
\end{equation}

During horizontal transit, encouraging operation near $V_{\mathrm{eco}}$ reduces the propulsion energy per unit horizontal distance so as to avoid both high-energy branches of the U-shaped curve, all while adaptively splitting the transmit power between downlink communication and sensing.

\subsection{Communication and Sensing Metrics}

\subsubsection{Communication Metric}
The achievable rate is adopted as the core communication measure. If the UAV transmits information to the $m$-th ground user in slot $t$, the downlink signal-to-interference-plus-noise ratio (SINR) is
\begin{equation}
\gamma_m(t)=\frac{p_m(t)|h_m(t)|^{2}}{\sum_{j\neq m}p_j(t)|h_j(t)|^{2}+\sigma^{2}},
\end{equation}
where $p_m(t)$ is the allocated power, $h_m(t)$ the channel gain, and $\sigma^{2}$ the noise power. The sequential-service rule determines when a node is credited as mission-served; it does not imply that non-served users are physically silent. During each slot, the available communication power is distributed among all users according to $\boldsymbol{\beta}(t)$, and the resulting aggregate throughput contributes to the system EE. By the Shannon formula, the instantaneous rate is $R_m(t)=B\log_2(1+\gamma_m(t))$~\cite{10251151} with $B$ the system bandwidth.

\subsubsection{Sensing Metric}
A rigorous CRB requires deriving the Fisher information matrix under a specific echo model, which would substantially increase the modeling and computational complexity of the sensing evaluation. Under fixed waveform and radar parameters, we therefore adopt a simplified accuracy surrogate that is inversely proportional to the echo signal-to-noise ratio (SNR)~\cite{10364735}. For a target $k$ at distance $d^{\mathrm{tar}}_{k}(t)$, the expected echo SNR is
\begin{equation}
\mathrm{SNR}_{\mathrm{echo},k}(t)=\frac{P_{\mathrm{sens}}(t)\,G_tG_r\lambda^{2}\sigma_{k}}{(4\pi)^{3}(d^{\mathrm{tar}}_{k}(t))^{4}\sigma_n^{2}},
\end{equation}
where $P_{\mathrm{sens}}(t)$ is the total sensing transmit power, $\sigma_{k}$ the radar cross section (RCS), and $\sigma_n^{2}$ the receiver noise power. Since the estimation-error lower bound is inversely proportional to the echo SNR, the sensing-error surrogate, denoted by $\epsilon_{k}^{\mathrm{sens}}(t)=\xi_{k}/\mathrm{SNR}_{\mathrm{echo},k}(t)$, leads to the constraint
\begin{equation}
\epsilon_{k}^{\mathrm{sens}}(t)=\frac{\xi_{k}(4\pi)^{3}(d^{\mathrm{tar}}_{k}(t))^{4}\sigma_n^{2}}{P_{\mathrm{sens}}(t)\,G_tG_r\lambda^{2}\sigma_{k}}\le\Gamma_{\mathrm{sens}},\quad\forall t,
\label{eq:sensing_error}
\end{equation}
in which $\xi_{k}$ is a proportionality constant related to the radar hardware. The metric decreases, i.e., the resolution improves, when the sensing power increases or the observation distance shrinks.

In post-disaster environments filled with debris, the radar cross section (RCS) $\sigma_k$ of trapped targets is typically weak and heavily attenuated. Consequently, maintaining a stringent sensing-error-surrogate threshold is important for ensuring reliable detection, explicitly forcing the UAV to physically approach the targets despite the aerodynamic energy cost.

\section{Problem Formulation}
\label{sec:problem}

%\subsection{Normalized Energy-Efficiency Objective}
The system EE is defined as the ratio of the total number of bits delivered to the ground users over the mission period to the total energy consumed by the UAV,
\begin{equation}
\eta_{\mathrm{EE}}(\mathbf{x})=\frac{\sum_{n=1}^{N}\sum_{m=1}^{M}R_m[n]\,\delta t}{\sum_{n=1}^{N}\mathcal{P}_{\mathrm{total}}[n]\,\delta t},
\label{eq:ee}
\end{equation}
where the variable set $\mathbf{x} = \{\mathbf{q}[n], \omega_r[n], \boldsymbol{\beta}[n]\}$ encompasses the 3D positions, the power-split ratios, and the power-allocation vectors across the trajectory. 

Because the magnitude of $\eta_{\mathrm{EE}}$ depends on the bandwidth and power settings and hence has no fixed range, we further define the \emph{normalized energy efficiency}
\begin{equation}
\hat{\eta}_{\mathrm{EE}}(\mathbf{x})=\frac{\eta_{\mathrm{EE}}(\mathbf{x})}{\eta_{\mathrm{EE}}^{\mathrm{ref}}},
\label{eq:ee_norm}
\end{equation}
where $\eta_{\mathrm{EE}}^{\mathrm{ref}}$ is a fixed reference scale (2.0~bits/J in this work) used for reward normalization. Since $\eta_{\mathrm{EE}}^{\mathrm{ref}}$ is a positive constant, \eqref{eq:ee_norm} is a monotonically increasing linear map that does not move the optimum, i.e.,
\begin{equation}
\argmax_{\mathbf{x}}\ \hat{\eta}_{\mathrm{EE}}(\mathbf{x})=\argmax_{\mathbf{x}}\ \eta_{\mathrm{EE}}(\mathbf{x}).
\end{equation}
Its magnitude no longer fluctuates with the system parameters, which improves cross-scenario comparability.

%\subsection{Constraints}
Guaranteeing the feasibility of the solution in practice requires satisfying communication, sensing, and kinematic constraints simultaneously. Let $\mathcal{T}_m^{\mathrm{serv}}$ and $\mathcal{T}_{k}^{\mathrm{serv}}$ denote the sets of slots at which user $m$ and target $k$ are actively served, respectively. The instantaneous rate of every user must meet a minimum requirement during its service slots,
\begin{equation}
R_m(t)\ge R_{\min},\quad\forall m,\; t\in\mathcal{T}_m^{\mathrm{serv}},
\end{equation}
and the simplified accuracy metric of the currently tracked target must not exceed the tolerable threshold as in~\eqref{eq:sensing_error},
\begin{equation}
\epsilon_{k}^{\mathrm{sens}}(t)\le\Gamma_{\mathrm{sens}},\quad\forall k,\; t\in\mathcal{T}_{k}^{\mathrm{serv}}.
\end{equation}
The horizontal and vertical speeds $\|\mathbf{v}_{xy}(t)\|$ and $|v_z(t)|$ are bounded by $V_{\max}$ and $V_{z,\max}$, the acceleration $\|\mathbf{a}(t)\|$ by $a_{\max}$, and the trajectory must satisfy the initial boundary condition $\mathbf{q}(0)=\mathbf{q}_{\mathrm{start}}$. Let $\mathcal{A}_{\mathrm{fly}}$ denote the prescribed 3D flight region, whose horizontal projection is the $1000\times1000$ m$^2$ mission area and whose altitude range is $[H_{\min},H_{\max}]$. The horizontal coordinates are constrained within the flight area such that $\mathbf{q}(t) \in \mathcal{A}_{\mathrm{fly}}$. Limited by the onboard radio-frequency hardware, the communication and sensing powers additionally satisfy $P_{\mathrm{comm}}(t)+P_{\mathrm{sens}}(t)\le P_{\max}$.

Beyond the instantaneous power budget, the mission is constrained by the onboard battery capacity. Over the whole mission the consumed energy must not exceed the strict energy budget $E_{\max}$,
\begin{equation}
\sum_{n=1}^{N}\mathcal{P}_{\mathrm{total}}[n]\,\delta t\le E_{\max}.
\label{eq:energy_budget}
\end{equation}
It is worth emphasizing that, because $\mathcal{P}_{\mathrm{total}}[n]$ varies nonlinearly with the flight speed through \eqref{eq:pfly}, the same flight duration consumes different amounts of energy at different speeds. We therefore adopt \emph{energy}, rather than time, as the criterion of this constraint.

In the implementation, the episode terminates if the accumulated energy exceeds the strict budget $E_{\max}$ or if the maximum step horizon $N_{\max}$ is reached. An episode that terminates without completing all visits is declared unsuccessful. The precise derivation of the nominal $E_{\max}$ capacity and its relationship with the horizontal flight power is further elaborated in Section~\ref{subsec:params}.

%\subsection{Problem Statement}
The optimization variables are the trajectory positions and radio-resource variables, with velocity and acceleration determined by the discrete-time kinematic model. Collecting the objective and the constraints, the joint optimization problem is formulated as
\begin{subequations}\label{eq:P0}
\begin{align}
(\mathrm{P}1)\colon\ &\max_{\{\mathbf{q}(t),\,\omega_r(t),\,\beta_m(t)\}}\ \hat{\eta}_{\mathrm{EE}}(\mathbf{x}) \\
\text{s.t.}\quad & R_m(t)\ge R_{\min},\quad\forall m,\;t\in\mathcal{T}_m^{\mathrm{serv}} \\
& \epsilon_{k}^{\mathrm{sens}}(t)\le\Gamma_{\mathrm{sens}},\quad\forall k,\;t\in\mathcal{T}_{k}^{\mathrm{serv}} \\
& \mathbf{q}(t+1)=\mathbf{q}(t)+\mathbf{v}(t)\delta t,\quad t=0,\ldots,N-1 \\
& H_{\min}\le z(t)\le H_{\max} \\
& \mathbf{q}(t)\in\mathcal{A}_{\mathrm{fly}} \\
& \|\mathbf{v}_{xy}(t)\|\le V_{\max},\quad |v_z(t)|\le V_{z,\max} \\
& \|\mathbf{a}(t)\|\le a_{\max} \\
& P_{\mathrm{comm}}(t)+P_{\mathrm{sens}}(t)\le P_{\max},\quad\forall t \\
& \sum_{n=1}^{N}\mathcal{P}_{\mathrm{total}}[n]\,\delta t\le E_{\max} \\
& \mathbf{q}(0)=\mathbf{q}_{\mathrm{start}}.
\end{align}
\end{subequations}
The RF power-budget constraint is automatically satisfied by the action parameterization. Since (P1) couples nonlinear trajectory-dependent channel conditions, propulsion energy, and resource allocation under sensing and QoS constraints, obtaining a globally optimized solution with low online computational cost is challenging. This motivates the DRL framework developed next.

\section{Curriculum-Guided SAC for Joint Trajectory and Resource Design}
\label{sec:method}

To obtain a low-complexity online policy for the continuous control problem in (P1), we formulate the problem as an MDP and adopt SAC as the underlying off-policy continuous-control algorithm. Built on the maximum-entropy framework, SAC maintains favorable convergence and robustness in complex continuous action spaces. Importantly, the proposed algorithm is not a plain application of vanilla SAC: on top of its maximum-entropy core we integrate two mechanisms tailored to the high-dimensional non-convex nature of the UAV-ISAC task, namely a curriculum-learning schedule and a normalized composite reward, which together turn it into a curriculum-guided joint optimization framework.

\subsection{MDP Formulation}
Problem (P1) is first cast as a Markov decision process (MDP).

\subsubsection{State Space}
The state $\mathbf{s}_t\in\mathcal{S}$ collects the principal observable features affecting EE, rate, and sensing accuracy,
\begin{equation}
\mathbf{s}_t=[\mathbf{q}(t),\,\mathbf{v}(t),\,\mathbf{d}_{\mathrm{CU}}(t),\,\mathbf{d}_{\mathrm{tar}}(t),\,\boldsymbol{\theta}_{\mathrm{CU}}(t),\,\mathbf{b}(t),\,\mathrm{Rem}_T],
\end{equation}
where $\mathbf{q}(t)$ and $\mathbf{v}(t)$ are the 3D position and instantaneous velocity, which jointly determine the displacement and the propulsion power; $\mathbf{d}_{\mathrm{CU}}(t)=[d^{\mathrm{CU}}_1(t),\dots,d^{\mathrm{CU}}_M(t)]$ and $\mathbf{d}_{\mathrm{tar}}(t)=[d^{\mathrm{tar}}_1(t),\dots,d^{\mathrm{tar}}_K(t)]$ are the distance vectors to the communication users and the sensing targets, respectively; $\boldsymbol{\theta}_{\mathrm{CU}}(t)=[\theta^{\mathrm{CU}}_1(t),\dots,\theta^{\mathrm{CU}}_M(t)]$ is the vector of elevation angles to the users; $\mathbf{b}(t)$ is the binary visited mask indicating the service status of each node; and $\mathrm{Rem}_T$ is the remaining mission time, which endows the agent with temporal awareness. The mission-tracked node is implicitly determined as the nearest unserved communication user or sensing target. When a sensing penalty is evaluated, $k^\star(t)$ is separately defined as the nearest unserved sensing target. It is worth noting that the state vector intentionally excludes explicit feedback on the remaining energy or battery capacity. Instead, the framework relies on temporal awareness ($\mathrm{Rem}_T$) coupled with the EE-centric reward structure. This design choice avoids explicitly conditioning the policy on a particular initial battery level. Its performance across different externally imposed energy budgets is evaluated empirically in Section~\ref{subsec:sensitivity}.

\subsubsection{Action Space}
The action $\mathbf{a}_t\in\mathcal{A}$ gathers all continuous decision variables,
\begin{equation}
\mathbf{a}_t=\bigl[v_x(t),\,v_y(t),\,v_z(t),\,\omega_r(t),\,\boldsymbol{\beta}(t)\bigr],
\end{equation}
where $v_x,v_y,v_z$ are the velocity commands along the three axes, $\omega_r(t)\in[0,1]$ is the power-split ratio between communication and sensing, and $\boldsymbol{\beta}(t)=[\beta_1(t),\dots,\beta_M(t)]$ is the relative power-allocation vector satisfying
\begin{equation}
\sum_{m=1}^{M}\beta_m(t)=1,\qquad \beta_m(t)\ge0,\ \forall m.
\end{equation}
The communication and sensing powers are thus $P_{\mathrm{comm}}(t)=\omega_r(t)P_{\max}$ and $P_{\mathrm{sens}}(t)=(1-\omega_r(t))P_{\max}$, and the power actually assigned to user $m$ is $p_m(t)=\beta_m(t)P_{\mathrm{comm}}(t)$. On the sensing side we do not allocate a dedicated power per target; instead a single aggregate sensing power $P_{\mathrm{sens}}(t)$ is applied to the currently tracked target. Because SAC operates on continuous actions, the velocity and split outputs are mapped through $\tanh(\cdot)$ to $[-1,1]$ and then linearly scaled, while $\boldsymbol{\beta}(t)$ is passed through a softmax so that the simplex constraint is strictly enforced. To enforce the kinematic acceleration constraint $\|\mathbf{a}(t)\| \le a_{\max}$ defined in (P1), the raw velocity commands $\mathbf{v}^{\mathrm{cmd}}(t)$ from the policy are passed through a first-order motion filter. Let $\Delta\mathbf{v}_t = \mathbf{v}_t^{\mathrm{cmd}} - \mathbf{v}_t$. The filtered velocity is given by
\begin{equation}
\mathbf{v}_{t+1} = \mathbf{v}_{t} + \min\!\left(1,\frac{a_{\max}\delta t}{\|\Delta\mathbf{v}_t\|}\right)\Delta\mathbf{v}_t,
\end{equation}
where the scaling factor is defined as $1$ when $\|\Delta\mathbf{v}_t\|=0$. This construction explicitly guarantees $\|\mathbf{v}_{t+1}-\mathbf{v}_{t}\|/\delta t \le a_{\max}$.

\subsection{Propulsion-Aware Composite Reward Shaping}
\label{subsec:reward}

To promote the mission-level EE objective while encouraging constraint satisfaction, we construct a composite reward. To provide step-wise feedback without requiring future trajectory information, we define an instantaneous EE surrogate $\eta_{\mathrm{inst}}(t) = \frac{\sum_{m=1}^{M}R_m(t)}{\mathcal{P}_{\mathrm{total}}(t)}$, and its normalized counterpart $\hat{\eta}_{\mathrm{inst}}(t) = \frac{\eta_{\mathrm{inst}}(t)}{\eta_{\mathrm{EE}}^{\mathrm{ref}}}$. Thus, at each step $t$ the agent receives
\begin{equation}
\begin{split}
r_t=\;&\lambda_{\mathrm{nav}}R_{\mathrm{nav}}(t)+\lambda_{\mathrm{visit}}R_{\mathrm{visit}}(t)+\lambda_{\mathrm{EE}}\,\hat{\eta}_{\mathrm{inst}}(t)+\lambda_{\mathrm{spd}}R_{\mathrm{spd}}(t)\\
&-w_{\mathrm{pen}}(e)\bigl(\mathrm{Penalty}_{\mathrm{rate}}(t)+\mathrm{Penalty}_{\mathrm{sens}}(t)\bigr)+R_{\mathrm{term}}(t),
\end{split}
\label{eq:reward}
\end{equation}
where the first line collects the positive guidance terms and the second the constraint penalties and the terminal settlement. The coefficients $\lambda_{\mathrm{nav}},\lambda_{\mathrm{visit}},\lambda_{\mathrm{EE}},\lambda_{\mathrm{spd}}>0$ balance the relative importance of the objectives. $\hat{\eta}_{\mathrm{inst}}(t)$ provides a step-wise proxy aligned with, but not identical to, the mission-level EE objective in \eqref{eq:ee_norm}, $w_{\mathrm{pen}}(e)$ is a penalty weight that grows with the curriculum progress, and $R_{\mathrm{term}}(t)$ acts only at the end of an episode. The navigation, node-visiting, EE, speed, and constraint-violation terms are normalized to $[0,1]$ or $[-1,1]$ before weighting, while the terminal settlement uses explicitly specified bonus and penalty magnitudes; the tuned values are listed in Table~\ref{tab:reward_weights}.

Each component has a distinct role in connecting the learning objective with the mission requirements. The instantaneous EE term provides a step-wise proxy for the core mission-level objective; the economic-speed reward $R_{\mathrm{spd}}$ steers the UAV toward $V_{\mathrm{eco}}$; the rate and sensing penalties incorporate the corresponding violations into the reward through penalty terms; the terminal settlement $R_{\mathrm{term}}$ encodes the boundary and completion constraints. Finally, the distance-progress navigation shaping $R_{\mathrm{nav}}$ provides dense directional feedback and is designed to encourage movement toward the nearest unserved node.

\subsubsection{Navigation Shaping}
To provide dense directional feedback in an otherwise sparse-reward environment, let $d_{\mathrm{near}}(t)$ be the distance from the UAV to the nearest unserved node. The distance-progress navigation shaping takes its per-step decrement, normalized by the maximum horizontal displacement in one slot,
\begin{equation}
R_{\mathrm{nav}}(t)=\frac{d_{\mathrm{near}}(t-1)-d_{\mathrm{near}}(t)}{V_{\max}\,\delta t},
\end{equation}
which is subsequently clipped to $[-1,1]$ before weighting.

\subsubsection{Node-Visiting Reward}
When node $j$ is effectively served for the first time at step $t$, i.e., the UAV enters its service range \emph{and} the corresponding rate or sensing requirement is met, a reward is granted,
\begin{equation}
R_{\mathrm{visit}}(t)=\sum_{j}\mathbf{1}\!\left[b_j(t-1)=0\ \wedge\ b_j(t)=1\right]\rho_j(t),
\end{equation}
where $b_j\in\{0,1\}$ is the visited mask of node $j$, $\mathbf{1}[\cdot]$ the indicator function, and $\rho_j(t)\in(0,1]$ a scaling factor increasing with the achieved service quality.

\subsubsection{Economic-Speed Reward}
To keep the UAV near the economic speed derived in \eqref{eq:veco}, we use
\begin{equation}
R_{\mathrm{spd}}(t)=1-\frac{\bigl|\,\|\mathbf{v}_{xy}(t)\|-V_{\mathrm{eco}}\,\bigr|}{V_{\max}},
\end{equation}
which is already normalized by $V_{\max}$ and encourages operation near the analytically derived economic speed.

\subsubsection{Constraint Penalties}
When the instantaneous rate of a user falls below $R_{\min}$, a penalty proportional to the relative rate deficit is imposed and averaged over the users so as to lie in $[0,1]$,
\begin{equation}
\mathrm{Penalty}_{\mathrm{rate}}(t)=\frac{1}{M}\sum_{m=1}^{M}\frac{\max\bigl(0,\,R_{\min}-R_m(t)\bigr)}{R_{\min}}\in[0,1].
\end{equation}
Although the hard communication constraint in (P1) is imposed only at service instants, the rate-violation penalty is evaluated at every step as a dense shaping signal to guide the policy toward favorable communication conditions.

For the sensing side, the penalty is continuously evaluated against the nearest unserved sensing target. Let $k^\star(t) = \arg\min_{k \in \mathcal{K}_{\mathrm{unserved}}} d^{\mathrm{tar}}_k(t)$ denote the currently tracked target. When all sensing targets have been served, we set $\mathrm{Penalty}_{\mathrm{sens}}(t)=0$. Otherwise, the sensing violation is measured in orders of magnitude beyond the threshold and squeezed into $[0,1]$. Let $\Delta_{\mathrm{sens}}(t) = \log_{10}\bigl(\epsilon_{k^\star(t)}^{\mathrm{sens}}(t)/\Gamma_{\mathrm{sens}}\bigr)$ denote the logarithmic sensing violation. The penalty is given by
\begin{equation}
\mathrm{Penalty}_{\mathrm{sens}}(t)=
\begin{cases}
\mathrm{clip}\!\left(\frac{\max\bigl(0,\Delta_{\mathrm{sens}}(t)\bigr)}{L_{\mathrm{sens}}},0,1\right), & \mathcal{K}_{\mathrm{unserved}} \neq \emptyset,\\
0, & \mathcal{K}_{\mathrm{unserved}} = \emptyset,
\end{cases}
\end{equation}
where $L_{\mathrm{sens}}$ is a normalization scale spanning roughly two decades. Since $\epsilon_{k^\star(t)}^{\mathrm{sens}}(t)$ depends jointly on the echo quality, the aggregate sensing power, and the UAV-target geometry, this penalty simultaneously drives the trajectory and the power-split policy.

\subsubsection{Terminal Settlement}
At the end of an episode we apply
\begin{equation}
R_{\mathrm{term}}=\begin{cases}
+\,r_{\mathrm{done}} + \lambda_{\mathrm{term}}\bar{\hat{\eta}}_{\mathrm{inst}}, & \text{all nodes served},\\[2pt]
-\,\zeta_{\mathrm{bnd}}, & \text{out of the service area},\\[2pt]
-\,\zeta_{\mathrm{unv}}\,N_{\mathrm{unv}}, & \text{otherwise},
\end{cases}
\end{equation}
where $N_{\mathrm{unv}}$ is the number of unserved nodes. Upon completion, an extra bonus proportional to the episode-average normalized EE is granted, defined as $\bar{\hat{\eta}}_{\mathrm{inst}} = \frac{1}{N_e}\sum_{t=1}^{N_e}\hat{\eta}_{\mathrm{inst}}(t)$ where $N_e$ is the duration of the episode, to further reinforce energy-saving behavior.

\begin{table}[!t]
\centering
\caption{Weights of the composite reward components}
\label{tab:reward_weights}
\footnotesize
\begin{tabular}{cll}
\hline
\textbf{Weight} & \textbf{Component} & \textbf{Value} \\
\hline
$\lambda_{\mathrm{nav}}$   & Navigation shaping reward        & $2.0$ \\
$\lambda_{\mathrm{visit}}$ & Node-visiting reward             & $150$ \\
$\lambda_{\mathrm{EE}}$    & Energy-efficiency reward         & $20$ \\
$\lambda_{\mathrm{spd}}$   & Economic-speed reward            & $50$ \\
$w_{\mathrm{pen}}(e)$      & Constraint penalty (curriculum)  & $0.5\to5.0$ \\
$r_{\mathrm{done}}$        & Mission-completion bonus         & $500$ \\
$\lambda_{\mathrm{term}}$  & Terminal EE bonus weight         & $300$ \\
$\zeta_{\mathrm{bnd}}$     & Out-of-boundary penalty          & $300$ \\
$\zeta_{\mathrm{unv}}$     & Unserved-node penalty (per node) & $80$ \\
\hline
\end{tabular}
\end{table}

\subsection{SAC Update and Curriculum Schedule}
Unlike conventional RL, which maximizes the cumulative reward alone, SAC incorporates the policy entropy $\mathcal{H}(\pi(\cdot|\mathbf{s}_n))$ into the objective~\cite{01290},
\begin{equation}
J(\pi)=\sum_{n=0}^{N_{\max}-1}\mathbb{E}_{(\mathbf{s}_n,\mathbf{a}_n)\sim\rho_\pi}\bigl[r_n+\alpha\mathcal{H}(\pi(\cdot|\mathbf{s}_n))\bigr],
\end{equation}
where the temperature $\alpha$ controls the trade-off between expected reward and policy entropy and is adjusted automatically during training~\cite{05905} to navigate the complex boundary formed by the communication and sensing constraints.

\begin{figure}[!t]
	\centering
	\includegraphics[width=.5\textwidth]{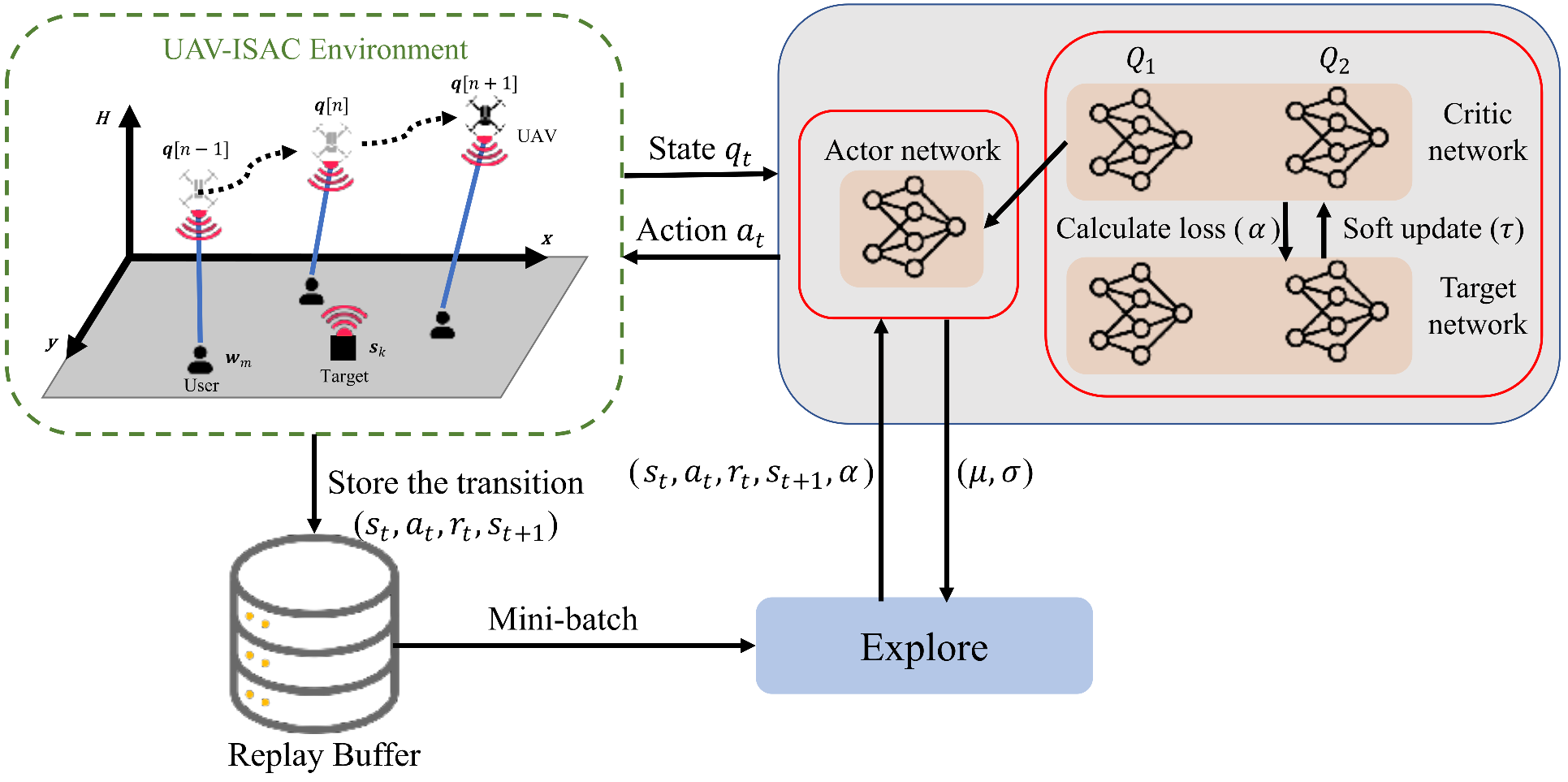}
	\caption{Architecture and update flow of the SAC agent.}
	\label{fig:sac}
\end{figure}

As shown in Fig.~\ref{fig:sac}, the architecture comprises an actor $\pi_\theta$ and twin critics $Q_{\phi_1},Q_{\phi_2}$. Given the limited onboard computing capability, the replay buffer and the network training reside on the ground edge server, and only the actor is kept onboard for online inference. Since the actor network consists of lightweight fully connected layers, the onboard forward-pass inference requires less than a millisecond per step, fully satisfying the real-time control requirements of highly dynamic UAV systems while effectively bypassing the onboard computational bottleneck. The critics are updated by minimizing the Bellman error
\begin{equation}
J_Q(\phi_c)=\mathbb{E}_{(\mathbf{s}_i,\mathbf{a}_i)\sim\mathcal{D}}\left[\tfrac{1}{2}\bigl(Q_{\phi_c}(\mathbf{s}_i,\mathbf{a}_i)-(r_i+\gamma\hat{V}(\mathbf{s}_{i+1}))\bigr)^{2}\right],
\end{equation}
and the actor by the reparameterization trick with
\begin{equation}
J_\pi(\theta)=\mathbb{E}_{\mathbf{s}_i\sim\mathcal{D}}\bigl[\alpha\log\pi_\theta(\mathbf{a}_i|\mathbf{s}_i)-Q_{\phi_{\min}}(\mathbf{s}_i,\mathbf{a}_i)\bigr].
\end{equation}

The second mechanism is the \emph{curriculum-learning schedule}, which follows the principle of presenting progressively harder training instances~\cite{bengio2009curriculum}. With a progress factor $c(e)=\min(e/E_c,1)$ over training episodes $e$, the minimum rate requirement $R_{\min}$, the sensing threshold $\Gamma_{\mathrm{sens}}$, and the proximity threshold $d_{\mathrm{th}}$ are tightened toward their final targets, while the violation penalty weight $w_{\mathrm{pen}}(e)$ is linearly amplified. 

However, relying solely on the standard automatic temperature adjustment of SAC is insufficient under a non-stationary curriculum. Under progressively tightened constraints, the automatically adapted entropy temperature may decrease substantially, reducing policy stochasticity during difficult curriculum transitions. To mitigate this effect, we implement a \emph{curriculum-synchronized entropy-temperature scheduling} mechanism by imposing an episode-dependent lower bound on $\alpha$, which decays stepwise in accordance with the curriculum progression (e.g., stepping down from $0.05$ during the early loose-constraint phases to $0.01$ in the final stringent phase). This heuristic mechanism helps preserve policy stochasticity during curriculum transitions.

The complete procedure is summarized in Algorithm~\ref{alg:cgsac}. Note that the curriculum and the synchronized annealing are treated strictly as external training-time scheduling mechanisms rather than part of the environmental state, ensuring the policy observes only the physically available system information. Because the curriculum parameters vary across training episodes, transitions in the replay buffer may be associated with different curriculum levels. We retain these transitions without retroactive reward recomputation; the finite FIFO buffer reduces the persistence of transitions generated under substantially earlier curriculum conditions.

\begin{algorithm}[!t]
\caption{Curriculum-Guided SAC for Joint UAV-ISAC Optimization}
\label{alg:cgsac}
\algsetup{linenosize=\small}
\small
\begin{algorithmic}[1]
\STATE \textbf{Initialize} actor $\pi_\theta$, critics $Q_{\phi_1},Q_{\phi_2}$, targets $\hat{Q}_{\phi_c}\!\leftarrow\!Q_{\phi_c}$, replay buffer $\mathcal{D}$, temperature $\alpha$
\STATE \textbf{Curriculum setup:} horizon $E_c$; ranges $R_{\min}^{0}\!\to\!R_{\min}^{\star}$, $\Gamma_{\mathrm{sens}}^{0}\!\to\!\Gamma_{\mathrm{sens}}^{\star}$, $d_{\mathrm{th}}^{0}\!\to\!d_{\mathrm{th}}^{\star}$, $w_{\mathrm{pen}}^{0}\!\to\!w_{\mathrm{pen}}^{\star}$, and exploration lower bound schedule $\alpha_{\min}: 0.05 \to 0.01$
\FOR{each episode $e=1,\dots,E$}
    \STATE $c\leftarrow\min(e/E_c,1)$
    \STATE $R_{\min}(e)\leftarrow\mathrm{clip}(c\,R_{\min}^{\star},R_{\min}^{0},R_{\min}^{\star})$;\ \ $\Gamma_{\mathrm{sens}}(e)\leftarrow10^{\,g_0-c(g_0-g_1)}$
    \STATE $d_{\mathrm{th}}(e)\leftarrow d_{\mathrm{th}}^{0}-c(d_{\mathrm{th}}^{0}-d_{\mathrm{th}}^{\star})$;\ \ $w_{\mathrm{pen}}(e)\leftarrow w_{\mathrm{pen}}^{0}+c(w_{\mathrm{pen}}^{\star}-w_{\mathrm{pen}}^{0})$
    \STATE Update the temperature lower bound $\alpha_{\min}$ based on the current curriculum phase
    \STATE Reset environment; initialize $\mathbf{s}_0$ and visited mask $\mathbf{b}\leftarrow\mathbf{0}$
    \FOR{each step $n=0,\dots,N_{\max}-1$}
        \STATE Sample $\mathbf{a}_n\sim\pi_\theta(\cdot|\mathbf{s}_n)$; execute to obtain $\mathbf{s}_{n+1}$, service metrics, and $\mathcal{P}_{\mathrm{total}}$
        \STATE Evaluate the composite reward $r_n$ in \eqref{eq:reward} with the current curriculum weights
        \STATE Update $\mathbf{b}$; store $(\mathbf{s}_n,\mathbf{a}_n,r_n,\mathbf{s}_{n+1})$ in $\mathcal{D}$
        \IF{update conditions are met}
            \STATE Sample a mini-batch from $\mathcal{D}$
            \STATE \textbf{Critics:} for each sampled transition $i$, $y_i=r_i+\gamma\bigl(\min_{c=1,2}\hat{Q}_{\phi_c}(\mathbf{s}_{i+1},\tilde{\mathbf{a}}_{i+1})-\alpha\log\pi_\theta(\tilde{\mathbf{a}}_{i+1}|\mathbf{s}_{i+1})\bigr)$; update $\phi_c$ by minimizing $\frac{1}{2}(Q_{\phi_c}(\mathbf{s}_i,\mathbf{a}_i)-y_i)^{2}$
            \STATE \textbf{Actor:} update $\theta$ by maximizing $\min_{c=1,2}Q_{\phi_c}(\mathbf{s}_i,\tilde{\mathbf{a}}_\theta)-\alpha\log\pi_\theta(\tilde{\mathbf{a}}_\theta|\mathbf{s}_i)$
            \STATE \textbf{Temperature and targets:} update $\alpha$ by minimizing $J(\alpha)$ subject to $\alpha \ge \alpha_{\min}$; soft update $\hat{\phi}_c\leftarrow\tau\phi_c+(1-\tau)\hat{\phi}_c$
        \ENDIF
    \ENDFOR
\ENDFOR
\end{algorithmic}
\end{algorithm}

\begin{table*}[!t]
\centering
\caption{Simulation parameters of the UAV-ISAC system and the SAC algorithm}
\label{tab:parameters}
\footnotesize
\setlength{\tabcolsep}{4pt}
\begin{tabular}{cll|cll}
\hline
\textbf{Symbol} & \textbf{Description} & \textbf{Value} & \textbf{Symbol} & \textbf{Description} & \textbf{Value} \\
\hline
$D_{\mathrm{area}}$  & Simulation area              & $1000\times1000$~m$^2$ & $B$                      & System bandwidth              & $10$~MHz \\
$H_{\min}$           & Minimum flight altitude      & $50$~m                 & $\sigma^{2}$             & Noise power                   & $10^{-11}$~W \\
$H_{\max}$           & Maximum flight altitude      & $150$~m                & $R_{\min}$               & Minimum rate requirement      & $5\to25$~Mbps \\
$M$                  & Number of communication users& $3$                    & $\Gamma_{\mathrm{sens}}$ & Sensing accuracy threshold    & $10^{12}\to10^{10}$ \\
$K$                  & Number of sensing targets    & $3$                    & $g_0, g_1$               & Sensing threshold log-exponents & $12, 10$ \\
$V_{\max}$           & Maximum horizontal speed     & $40$~m/s               & $d_{\mathrm{th}}^0, d_{\mathrm{th}}^\star$ & Initial/final proximity thresholds & $80, 60$~m \\
$V_{z,\max}$         & Maximum vertical speed       & $20$~m/s               & $E_c$                    & Curriculum horizon (episodes) & $3000$ \\
$P_{\mathrm{hover}}$ & Hovering power               & $80$~W                 & $\alpha_{\min}$          & Entropy temp. lower bound     & $0.05\to0.01$ \\
$P_{\mathrm{full}}$  & Full-speed flight power      & $250$~W                & $\eta$                   & Learning rate                 & $3\times10^{-4}$ \\
$P_{\mathrm{elec}}$  & Hardware circuit power       & $10$~W                 & $\gamma$                 & Discount factor               & $0.99$ \\
$P_{\max}$           & Maximum transmit power budget& $5$~W                  & $\tau$                   & Soft-update coefficient       & $0.005$ \\
$\kappa_z$           & Vertical climbing power coef.& $12.0$~J/m             & $B_{\mathrm{size}}$      & Batch size                    & $512$ \\
$E$                  & Total training episodes      & $5000$                 & $N_{\max}$               & Maximum steps per episode     & $800$ \\
$\delta t$           & Time slot duration           & $1$~s                  & $L_{\mathrm{sens}}$      & Sensing penalty norm. scale   & $2$ \\
\hline
\end{tabular}
%\vspace{-1em}
\end{table*}

\subsection{Computational Complexity Analysis}
To evaluate the practical feasibility of the proposed CG-SAC framework, we analyze its computational and space complexities. The architecture relies on an asymmetric edge-assisted deployment: training is offloaded to the ground server, while only the actor network operates onboard the UAV.

Let $|\mathcal{S}|$ and $|\mathcal{A}|$ denote the dimensions of the state and action spaces, respectively. Suppose the actor and critic networks are implemented as \textit{multi-layer perceptrons} (MLPs) with $H$ hidden layers, each containing $U$ neurons. 

During the \emph{training phase} at the ground edge server, the time complexity per update step is dominated by the forward and backward propagation through one actor and two critic networks, alongside the target network soft updates. The time complexity is bounded by $\mathcal{O}(|\mathcal{S}|U + (H-1)U^2 + U|\mathcal{A}|)$ per network. The space complexity is primarily dictated by the experience replay buffer $\mathcal{D}$ with capacity $B_{\mathrm{size}}$, requiring $\mathcal{O}(B_{\mathrm{size}}(|\mathcal{S}| + |\mathcal{A}|))$ memory. These requirements are easily handled by standard ground servers.

During the \emph{online inference phase} onboard the UAV, the critics and the replay buffer are entirely discarded. During inference, the mean action of the trained actor is used for deterministic control. Thus, the onboard time complexity per control step is $\mathcal{O}(|\mathcal{S}|U + (H-1)U^2 + U|\mathcal{A}|)$, and the space complexity is reduced to storing the actor network weights, which is also $\mathcal{O}(|\mathcal{S}|U + (H-1)U^2 + U|\mathcal{A}|)$. On our test platform, the measured actor inference latency is below 1 ms per control step. For the adopted 2-layer, 256-unit multi-layer perceptron architecture, the resulting actor inference requires only a small number of matrix operations. The measured latency indicates that the trained actor can support sub-millisecond control updates on the tested platform.

\section{Performance Evaluation}
\label{sec:simulation}

\subsection{Simulation Setup}
\label{subsec:params}
A 3D simulation environment is built with $3$ communication users and $3$ sensing targets randomly distributed in the area, corresponding to the small relief unit described in Section~\ref{sec:system}. The flight altitude is confined to $50\sim150$~m and the system bandwidth is $10$~MHz. Each slot lasts $\delta t=1$~s and an episode is capped at $N_{\max}=800$ slots. 

From the energy model of Section~\ref{subsec:energy}, the nominal upper bound of the instantaneous total power during horizontal flight is $\mathcal{P}_{\mathrm{total}}^{\max} \approx 265$~W, attained when flying at $V_{\max}$. The onboard energy budget is strictly set to $E_{\max}=212$~kJ ($\approx58.9$~Wh). This value is established as a nominal energy-budget baseline corresponding to the energy required to fly at the maximum horizontal speed $V_{\max}$ for $N_{\max}=800$ slots without climbing (i.e., $\approx 265\text{ W} \times 800\text{ s}$). While vertical climbing incurs additional instantaneous power according to \eqref{eq:pfly}, the overall energy budget remains hard-capped at $212$~kJ, forcing the agent to manage its 3D maneuvers efficiently to avoid premature energy exhaustion. An episode that reaches $N_{\max}$ slots or exceeds $E_{\max}$ without completing all visits is declared unsuccessful. To improve exploration in the non-convex space, a curriculum tightens the minimum rate to $25$~Mbps, the sensing threshold to $10^{10}$, and the proximity threshold to $60$~m, while amplifying the penalty weight and the EE reward. All parameters are listed in Table~\ref{tab:parameters}.

For performance comparison, we evaluate CG-SAC against four continuous-control DRL baselines: DDPG~\cite{lillicrap2016ddpg}, TD3~\cite{fujimoto2018td3}, PPO~\cite{schulman2017ppo}, and the standard maximum-entropy Vanilla SAC~\cite{01290}. Vanilla SAC serves as a baseline for evaluating the combined effect of curriculum guidance and synchronized entropy scheduling in this highly constrained environment. All non-curriculum baselines use the fixed final task constraints, whereas curriculum-enabled methods follow identical prescribed schedules and episode budgets to ensure fairness.

The subsequent performance evaluation is systematically structured into four parts. First, we analyze the training convergence and policy stability against the baseline algorithms. Second, the macro-level inference performance is evaluated across randomized test scenarios. Third, a micro-level analysis investigates the service quality distributions and parameter sensitivity. Finally, an ablation study examines the roles of four selected reward components (including shaping terms and penalties).

\subsection{Convergence and Baseline Comparison}
\begin{figure}[!t]
    \centering
    \subfloat[Task completion stability.]{\includegraphics[width=0.48\columnwidth]{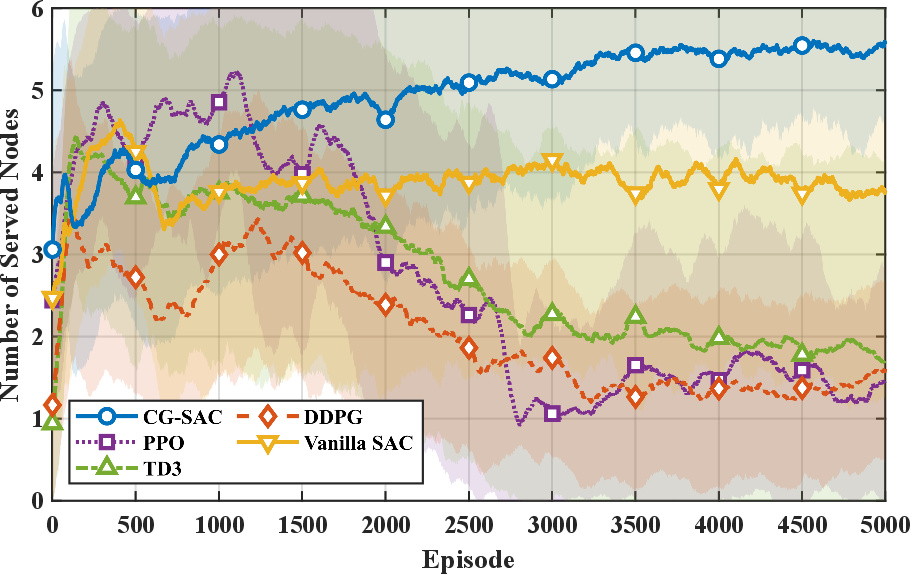}%
    \label{fig:convergence:visited}}
    \hfil
    \subfloat[Energy efficiency trend.]{\includegraphics[width=0.485\columnwidth]{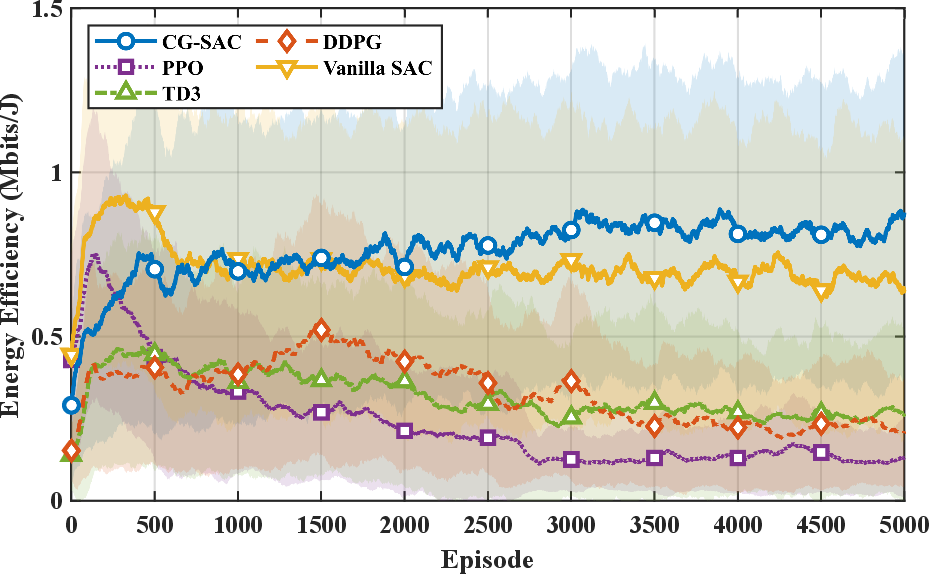}%
    \label{fig:convergence:ee}}\\

    \subfloat[Path efficiency trend.]{\includegraphics[width=0.49\columnwidth]{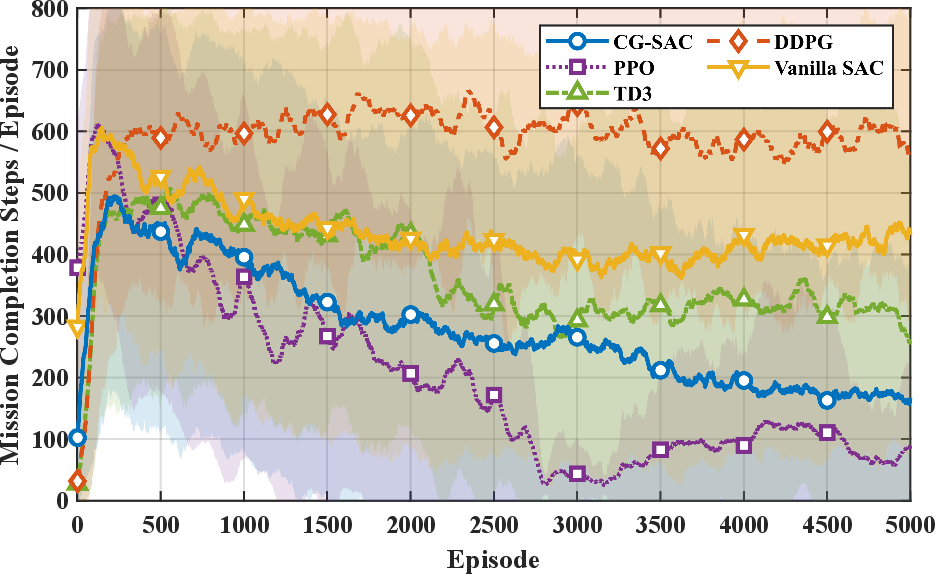}%
    \label{fig:convergence:steps}}
    \hfil
    \subfloat[Episode-average horizontal speed.]{\includegraphics[width=0.485\columnwidth]{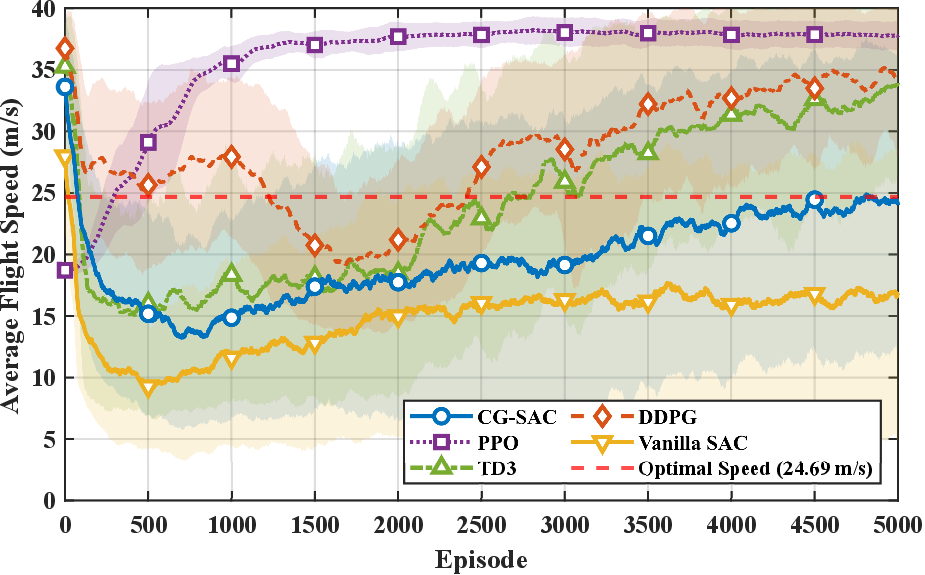}%
    \label{fig:convergence:speed}}
    \caption{Training convergence behavior and physical state adaptation of the proposed CG-SAC and baseline algorithms. (a) Number of served nodes. (b) Energy efficiency. (c) Mission completion steps per episode. (d) Episode-average horizontal flight speed and the model-derived propulsion-economic speed reference.}
    \label{fig:convergence}
\end{figure}

Fig.~\ref{fig:convergence} presents a holistic view of the training trajectories over 5000 training episodes across four key metrics, merging the behavioral analysis of the proposed method and the baselines.

As shown in Fig.~\ref{fig:convergence:visited}, the average number of served nodes rises steadily and approaches the maximum of six nodes per episode. Among the baselines, Vanilla SAC demonstrates the most resilience, plateauing at approximately 4 nodes per episode. Without the proposed curriculum and synchronized entropy-temperature scheduling, Vanilla SAC converges to a conservative sub-optimal operating regime. In contrast, PPO and TD3 initially show moderate coverage but suffer severe policy collapses after episode 1500, dropping to approximately 1 to 2 nodes, while DDPG exhibits persistently poor coverage throughout the training.

These behavioral patterns are directly reflected in the energy efficiency trends depicted in Fig.~\ref{fig:convergence:ee}. The EE of CG-SAC robustly climbs and stabilizes around 0.8 to 0.9 Mbits/J. Vanilla SAC settles into a sub-optimal plateau around 0.6 to 0.7 Mbits/J, establishing itself as the strongest baseline but still falling significantly short of the proposed method. Meanwhile, the EE of TD3 declines in tandem with its completion rate, and PPO drops to a low plateau near 0.1 to 0.2 Mbits/J, demonstrating the inability of these standard methods to maintain efficient flight profiles under strict constraints.

The underlying causes of these performance gaps are fundamentally illustrated by the path efficiency in Fig.~\ref{fig:convergence:steps} and the flight speed adaptation in Fig.~\ref{fig:convergence:speed}. CG-SAC progressively reduces the episode length toward approximately 200 steps during training, while the fully trained policy achieves 107.6 steps on the unseen test scenarios. Vanilla SAC stagnates at approximately 400 steps, whereas TD3 and DDPG exhibit substantially longer and more unstable trajectories, often approaching the episode limit. Furthermore, Fig.~\ref{fig:convergence:speed} reveals that under the explicit economic-speed shaping term, CG-SAC learns a policy that closely aligns its flight speed with the model-derived propulsion-economic speed reference of 24.69 m/s. Vanilla SAC exhibits a more conservative speed profile, which prolongs mission duration and is consistent with its lower EE. PPO adopts a high-consumption strategy of sprinting near the maximum speed (nearly 38 m/s), which severely penalizes its propulsion energy and leads to frequent early terminations, resulting in fewer steps per episode despite its poor mission completion.

\subsection{Inference Performance Evaluation}

\subsubsection{Macro-Level Performance Evaluation}
\begin{figure}[!t]
    \centering
    \subfloat[Distribution of average EE.]{\includegraphics[width=0.485\columnwidth]{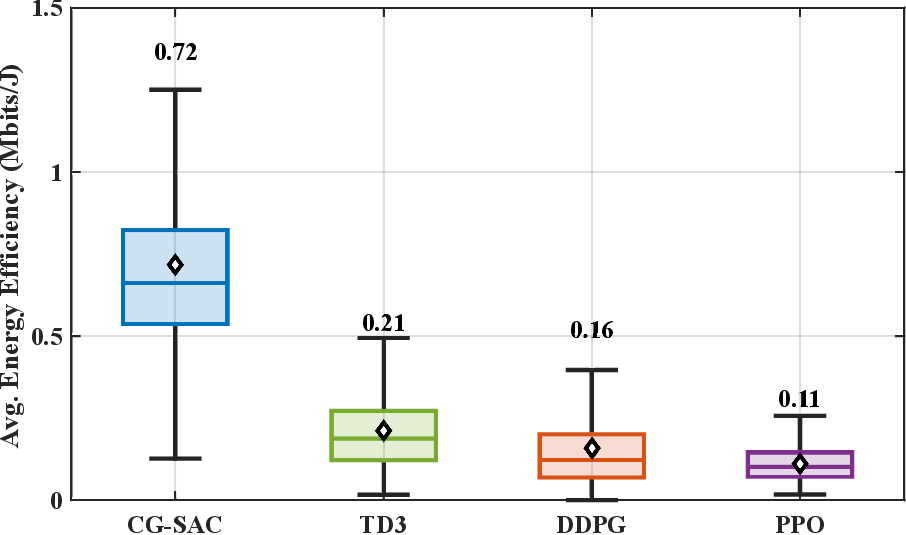}%
    \label{fig:macro_analysis:ee}}
    \hfil
    \subfloat[Average mission completion steps.]{\includegraphics[width=0.49\columnwidth]{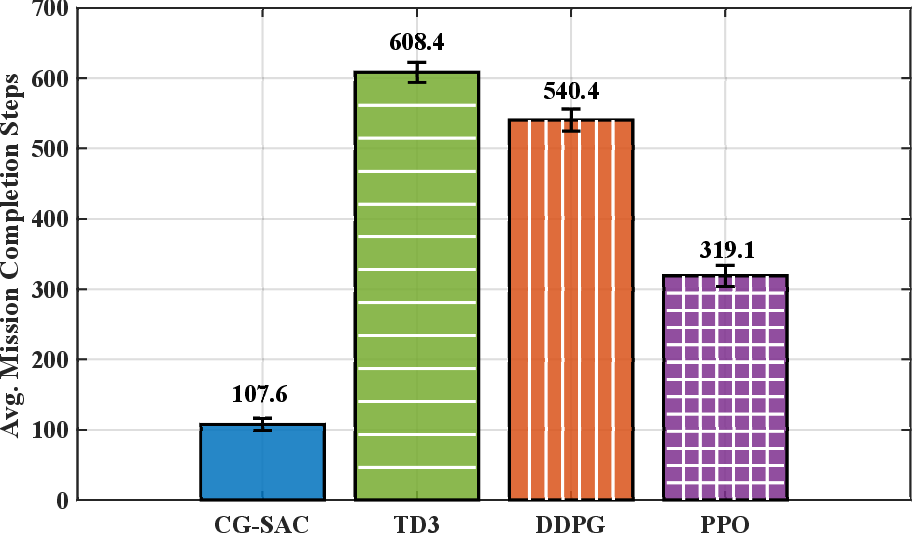}%
    \label{fig:macro_analysis:steps}}\\    
    \subfloat[Communication satisfaction rate.]{\includegraphics[width=0.485\columnwidth]{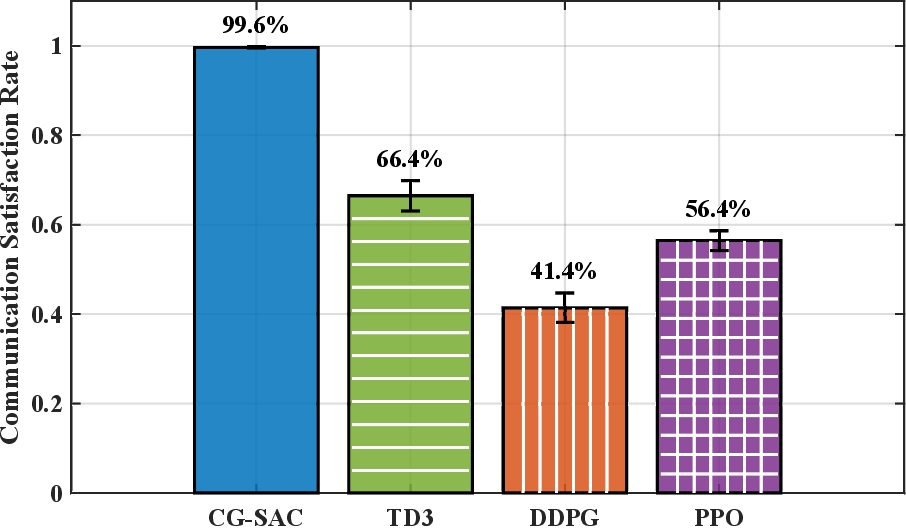}%
    \label{fig:macro_analysis:comm}}
    \hfil
    \subfloat[Sensing satisfaction rate.]{\includegraphics[width=0.485\columnwidth]{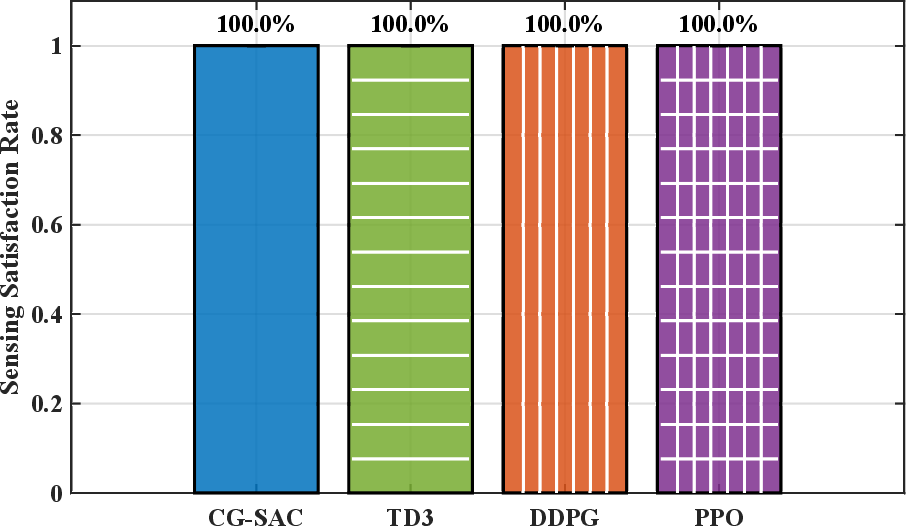}%
    \label{fig:macro_analysis:sens}}
    \caption{Macro-level performance statistics of the proposed CG-SAC and baseline algorithms. (a) Distribution of average energy efficiency. (b) Average mission completion steps. (c) Communication satisfaction rate at visit instants ($R_{\rm min} \ge 25$ Mbps). (d) Sensing satisfaction rate at visit instants ($\epsilon^{\mathrm{sens}} \le 10^{10}$).}
    \label{fig:macro_analysis}
\end{figure}

To validate generalization, the trained policies are deployed across $2000$ randomized test scenarios, with macro-level statistics summarized in Fig.~\ref{fig:macro_analysis}.

In Fig.~\ref{fig:macro_analysis:ee}, the proposed CG-SAC demonstrates a substantial advantage in energy efficiency, achieving an average EE of $0.72$~Mbits/J. This performance is over three times higher than the best baseline, TD3, which only yields $0.21$~Mbits/J. The boxplot distribution reveals that CG-SAC flexibly adapts its energy expenditure to various topological layouts, whereas the baseline policies exhibit less efficient flight behaviors across the tested layouts. This difference is further reflected by the mission completion steps in Fig.~\ref{fig:macro_analysis:steps}. For each method, the average mission completion steps are computed exclusively over successfully completed missions. CG-SAC requires an average of $107.6$ mission steps over successfully completed episodes. Conversely, DDPG and TD3 suffer from severe navigational wandering, consuming over $500$ steps, while PPO requires $319.1$ steps, leading to substantially higher propulsion-energy consumption. Under the final stringent constraints, Vanilla SAC exhibits an insufficient full-mission completion rate; therefore, its successful-mission-conditioned metrics are reported as N/A in the subsequent macro- and micro-level analyses. Its role in the evaluation is primarily to isolate the combined contribution of curriculum guidance and synchronized entropy-temperature scheduling in the training convergence analysis.

Beyond navigational efficiency, satisfying the QoS requirement at the service instants is a key mission objective. Fig.~\ref{fig:macro_analysis:comm} evaluates the communication satisfaction rate, calculated exclusively at the visit instants to reflect the true service reliability. CG-SAC successfully satisfies the strict rate threshold ($R_{\min} \ge 25$~Mbps) in $99.6\%$ of the service events. The baselines are less effective at adapting their 3D positions to favorable channel conditions, resulting in severe degradation. TD3 achieves $66.4\%$, PPO achieves $56.4\%$, and DDPG fails in more than half of its attempts with a mere $41.4\%$ success rate. This contrast highlights that navigating to a two-dimensional coordinate alone is insufficient in this setting, because the achievable service quality also depends on the UAV altitude and flight state; the agent must actively manage its three-dimensional altitude and speed to optimize the dynamic channel capacity.

Finally, Fig.~\ref{fig:macro_analysis:sens} indicates that all tested policies satisfy the basic sensing-feasibility criterion at service instants ($\epsilon^{\mathrm{sens}} \le 10^{10}$, where time and target indices are omitted for brevity). This uniform success is largely attributable to the spatial distance ceiling enforced during the state transitions, which naturally guarantees a baseline level of radar return power once a visit is triggered. However, simply crossing this conservative threshold does not imply identical sensing precision, a nuance that necessitates the subsequent micro-level distribution analysis.

\subsubsection{Micro-Level Service Quality and Sensitivity Analysis}

\begin{figure}[!t]
    \centering
    \subfloat[CDF of served user rate.]{\includegraphics[width=0.49\columnwidth]{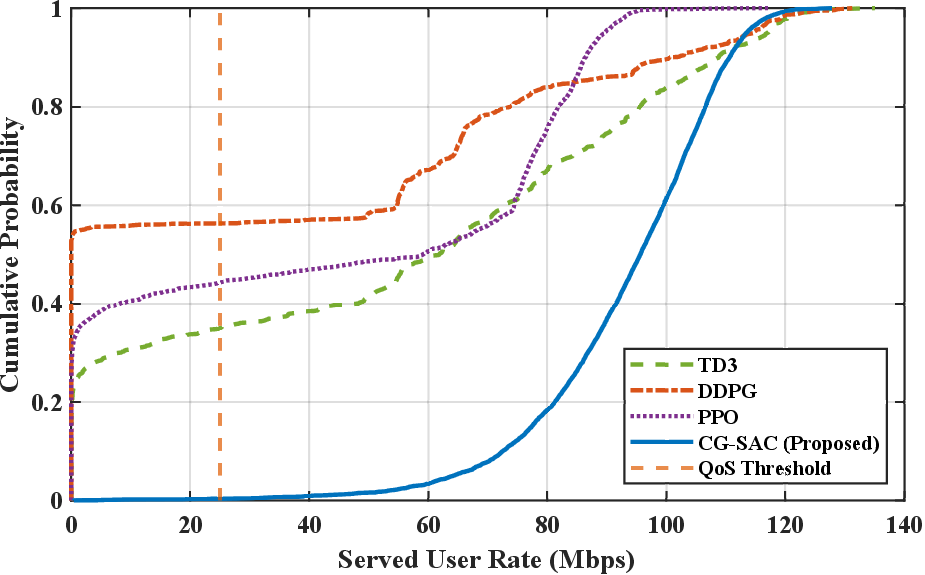}%
    \label{fig:micro_analysis:comm_cdf}}
    \hfil
    \subfloat[CDF of sensing-error surrogate.]{\includegraphics[width=0.49\columnwidth]{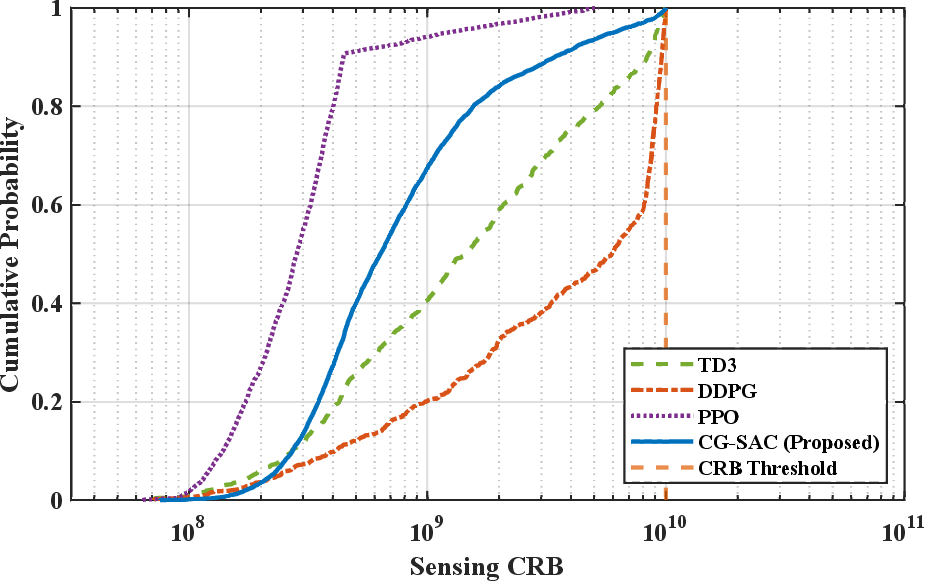}%
    \label{fig:micro_analysis:sens_cdf}}\\    
    \subfloat[Rate threshold sensitivity.]{\includegraphics[width=0.49\columnwidth]{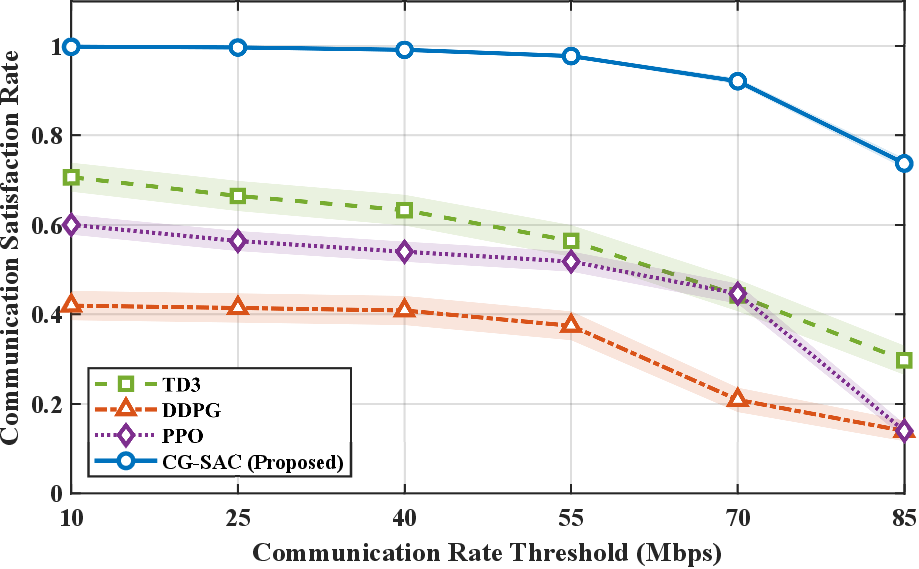}%
    \label{fig:micro_analysis:comm_sens}}
    \hfil
    \subfloat[Sensing channel sensitivity.]{\includegraphics[width=0.49\columnwidth]{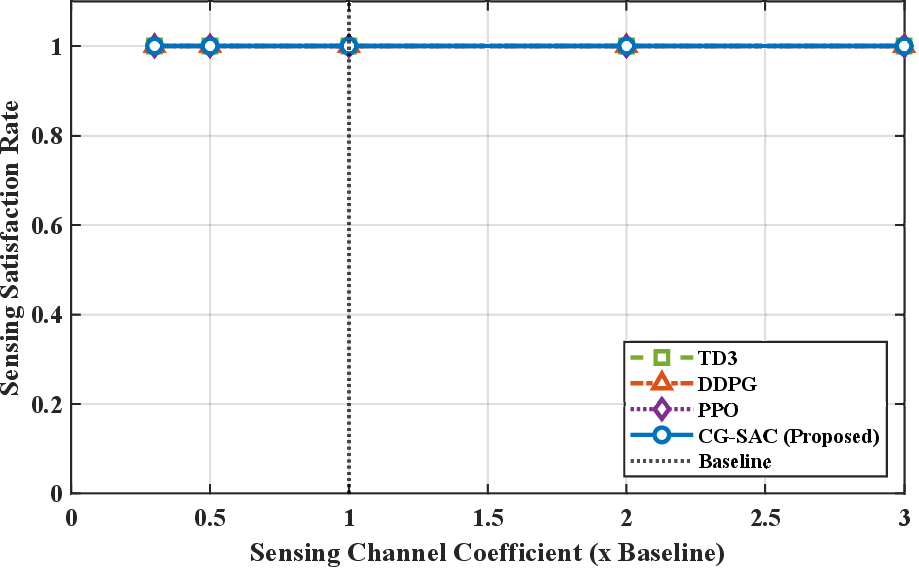}%
    \label{fig:micro_analysis:sens_robust}}
    \caption{Micro-level service quality and sensitivity evaluation of the proposed CG-SAC and baseline algorithms. (a) Cumulative distribution function of the served communication rate across all service attempts. (b) Cumulative distribution function of the sensing-error surrogate at visit instants. (c) Communication satisfaction probability under varying data rate thresholds. (d) Sensing satisfaction probability under varying sensing channel coefficients.}
    \label{fig:micro_analysis}
\end{figure}

To investigate the fine-grained performance and sensitivity to sensing-channel variations, we conduct a micro-level analysis focusing on the cumulative distributions and sensitivity behaviors depicted in Fig.~\ref{fig:micro_analysis}.

In Fig.~\ref{fig:micro_analysis:comm_cdf}, the \textit{cumulative distribution function} (CDF) of the served user rate reveals the distinct performance limitations of the baseline algorithms. While CG-SAC maintains a sharp, right-shifted distribution where nearly all visit-instant rates exceed the $25$~Mbps QoS threshold, the baselines exhibit substantial probability mass near zero rate. Specifically, DDPG and PPO experience immediate probability jumps of over $55\%$ and $35\%$ near zero, indicating that more than half of their attempted service interactions fail to establish meaningful communication links due to poor spatial alignment.

A complementary microscopic perspective is illustrated by the distribution of the sensing-error surrogate in Fig.~\ref{fig:micro_analysis:sens_cdf}. Although all algorithms manage to stay below the basic threshold of $10^{10}$, PPO and CG-SAC achieve substantially lower surrogate values across the entire distribution compared to TD3 and DDPG. While PPO also attains low surrogate values, it does so at the expense of communication performance (as evidenced by its high failure rate in Fig.~\ref{fig:micro_analysis:comm_cdf}), whereas CG-SAC effectively balances both objectives. This indicates that CG-SAC not only fulfills the loose constraint but actively optimizes the geometric observation angle and proximity without compromising ground user service quality.

Furthermore, Fig.~\ref{fig:micro_analysis:comm_sens} examines the communication satisfaction probability under stringent rate threshold sweeps ranging from $10$ to $85$~Mbps. As the requirement tightens, CG-SAC exhibits remarkable stability, preserving a satisfaction rate above $95\%$ up to $55$~Mbps and retaining over $73\%$ even at an extreme threshold of $85$~Mbps. In contrast, the baseline algorithms suffer rapid performance degradation, dropping below $50\%$ as soon as the threshold exceeds $55$~Mbps.

Finally, Fig.~\ref{fig:micro_analysis:sens_robust} examines the sensitivity of the trained policies to variations in the sensing-channel coefficient by scaling $\zeta_{\mathrm{sens}}$ from $0.3\times$ to $3.0\times$ of its baseline value. All algorithms maintain stable sensing satisfaction across the tested range, indicating that the sensing-feasibility metric is relatively insensitive to these channel-coefficient variations in the considered setting.

\subsection{Ablation Study of the Reward Design}
In a single-UAV ISAC system, the agent cannot distribute tasks among multiple units; it must sequentially satisfy stringent spatial, temporal, and communication constraints within a strictly limited onboard energy budget. To verify how the proposed reward framework guides the single UAV to untangle these highly coupled constraints, we conduct an ablation study to examine the roles of four selected reward shaping and penalty components: (i) \textit{w/o Nav} removes the navigation shaping $R_{\mathrm{nav}}$; (ii) \textit{w/o Rate} removes the rate penalty; (iii) \textit{w/o Sens} removes the sensing penalty; and (iv) \textit{w/o EE} removes the EE reward. The network architectures, hyper-parameters, and curriculum schedules remain identical across all variants.

Because the ablation variants are evaluated on a separate set of 100 unseen scenarios, the performance of the full model in Table~\ref{tab:ablation} is not directly identical to the 2000-scenario main evaluation. The evaluation metrics are summarized in Table~\ref{tab:ablation}. The completion rate represents the fraction of episodes where all six nodes are sequentially served. The rate and sensing satisfaction ratios are measured exclusively at the service instants, with the latter strictly evaluated under the $\epsilon^{\mathrm{sens}}\le 10^{10}$ criterion.

The results in Table~\ref{tab:ablation} reveal two distinct categories of performance degradation: a total collapse of mission feasibility (w/o Nav and w/o Sens) and a severe decline in energy efficiency (w/o Rate and w/o EE).

\begin{table*}[!t]
\centering
\caption{Test-stage performance of the reward-ablation variants over $100$ unseen scenarios}
\label{tab:ablation}
\footnotesize
\setlength{\tabcolsep}{4pt}
\begin{tabular}{lccccc}
\hline
\textbf{Variant} & \textbf{Completion (\%)} & \textbf{Rate Satisfaction (\%)} & \textbf{Sensing Satisfaction (\%)} & \textbf{Average Steps} & \textbf{Average EE (Mbits/J)} \\
\hline
\textbf{Full (proposed)}       & $91.0$ & $99.7$ & $89.9$ & $\mathbf{97.4}$ & $\mathbf{0.739}$ \\
w/o Rate (no rate penalty)     & $95.0$ & $99.3$ & $93.3$ & $39.1$ & $0.613$ \\
w/o Sens (no sensing penalty)   & $14.0$ & $97.8$ & $59.8$ & $597.8$ & $0.664$ \\
w/o EE (no EE reward)          & $91.0$ & $99.3$ & $93.5$ & $77.8$ & $0.628$ \\
w/o Nav (no navigation shaping)& $0.0$  & $98.3$ & $86.2$ & $647.4$ & $0.675$ \\
\hline
\end{tabular}
%\vspace{-1em}
\end{table*}

\subsubsection{Foundations of Mission Feasibility in 3D Space}
For a single UAV tasked with sequentially visiting multiple dispersed nodes, the continuous 3D state space is immense, and the terminal completion reward is inherently sparse. The w/o Nav variant completely fails to establish an effective routing behavior, yielding a $0.0\%$ completion rate and wandering for an average of $647.4$ steps. As illustrated in Fig.~\ref{fig:ablation_analysis:convergence}, without the dense gradient field provided by the distance-progress navigation shaping, the agent cannot learn the basic geometric concept of approaching a target. 

Conversely, the w/o Sens variant exhibits a progressive structural failure. Its task coverage initially rises but sharply degrades as the curriculum tightens. In an ISAC context, radar sensing is highly sensitive to the 3D relative distance and observation angle. Without the explicit sensing penalty, the single UAV optimizes its trajectory solely for proximity or communication, failing to satisfy the sensing-accuracy requirement represented by the surrogate metric. Consequently, its completion rate plummets to $14.0\%$, and its sensing satisfaction ratio degrades to $59.8\%$ (Table~\ref{tab:ablation}), suggesting that spatial navigation alone is insufficient to guarantee sensing feasibility. This behavior is corroborated by Fig.~\ref{fig:ablation_analysis:breakdown}, where w/o Sens demonstrates severe policy instability, with episode completions randomly distributed across the $1\sim3$ ($49\%$) and $4\sim5$ ($33\%$) ranges.

\subsubsection{Energy Efficiency and the Aerodynamic Dilemma}
The variants w/o Rate and w/o EE successfully complete the mission with high probabilities ($95.0\%$ and $91.0\%$), demonstrating that they are capable of navigating the topology. However, their fundamental flaw lies in energy blindness, which is particularly detrimental under a strict single-UAV energy budget.

\begin{figure}[!t]
    \centering
    \subfloat[Task completion stability.]{\includegraphics[width=0.47\columnwidth]{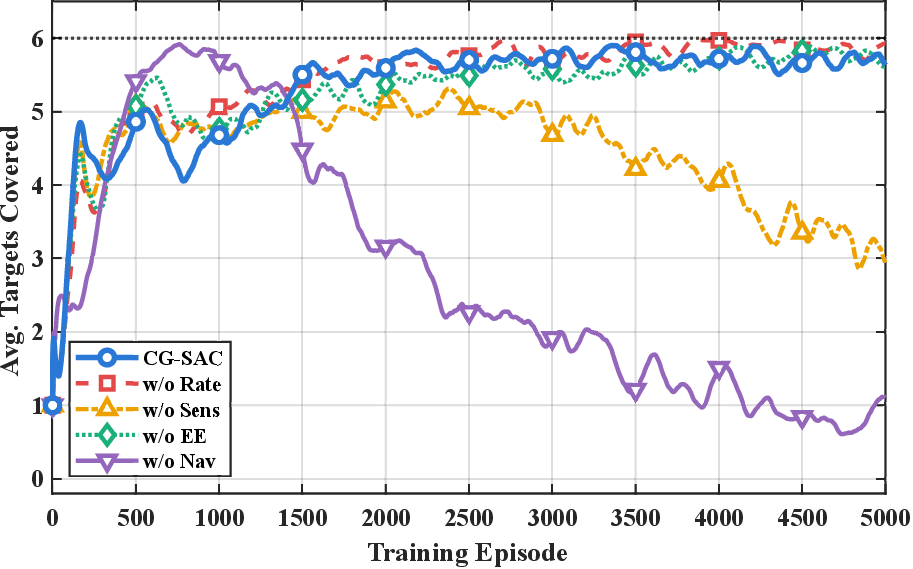}%
    \label{fig:ablation_analysis:convergence}}
    \hfil
    \subfloat[Average energy efficiency.]{\includegraphics[width=0.49\columnwidth]{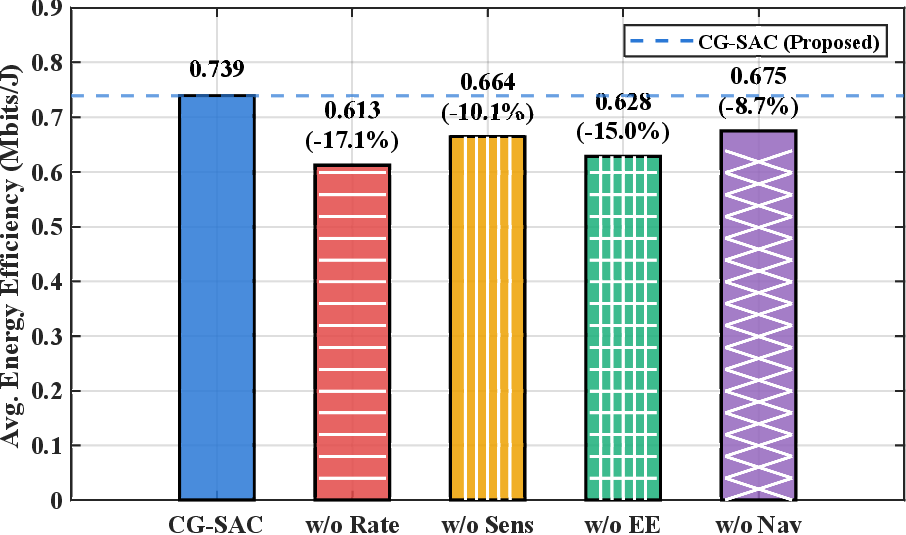}%
    \label{fig:ablation_analysis:ee}}\\    
    \subfloat[Flight speed adaptation.]{\includegraphics[width=0.485\columnwidth]{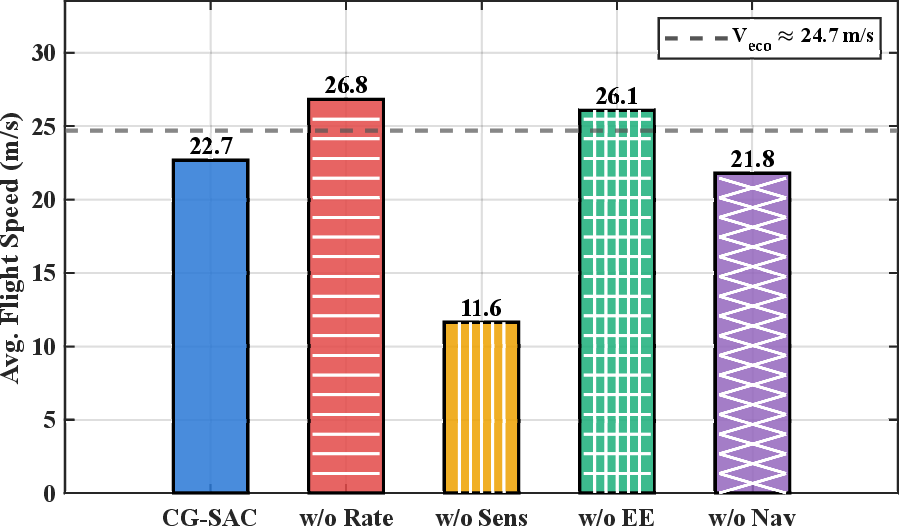}%
    \label{fig:ablation_analysis:speed}}
    \hfil
    \subfloat[Task completion breakdown.]{\includegraphics[width=0.49\columnwidth]{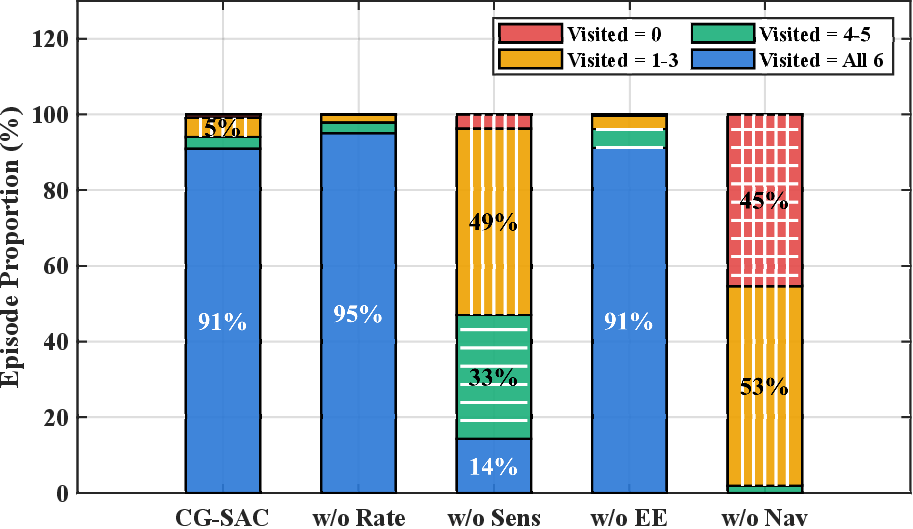}%
    \label{fig:ablation_analysis:breakdown}}
    \caption{Ablation study of the proposed CG-SAC framework under different reward components and architectural variants. (a) Training convergence trajectories of average targets covered. (b) Average energy efficiency and performance degradation relative to the full model. (c) Average flight speed compared to the model-derived propulsion-economic speed reference $V_{\mathrm{eco}}$. (d) Percentage breakdown of target completion proportions per episode across variants.}
    \label{fig:ablation_analysis}
\end{figure}

As shown in Fig.~\ref{fig:ablation_analysis:ee}, compared to the Full model ($0.739$~Mbits/J), the average EE drops significantly by $17.1\%$ for w/o Rate ($0.613$~Mbits/J) and $15.0\%$ for w/o EE ($0.628$~Mbits/J). The mechanical root of this inefficiency is captured in Fig.~\ref{fig:ablation_analysis:speed}. The proposed Full model yields an episode-average horizontal speed of $22.7$~m/s, operating efficiently just below the model-derived propulsion-economic speed reference of $24.7$~m/s. In contrast, the w/o Rate and w/o EE variants aggressively accelerate to $26.8$~m/s and $26.1$~m/s, respectively. 

This reveals the core physical dilemma of single-UAV control: the trade-off between mission duration and aerodynamic power consumption. Without the rate penalty, the UAV is not heavily penalized for establishing brief, low-quality communication links; thus, it adopts a ``sprinting'' strategy to rapidly finish the mission (averaging merely $39.1$ steps). Similarly, without the explicit EE reward, standard reinforcement learning simply maximizes the unnormalized step-wise returns by finishing as quickly as possible. Because the propulsion power of a rotary-wing UAV grows with the cube of its horizontal speed ($V^3$), sprinting leads to substantially higher propulsion-energy consumption. The full CG-SAC model explicitly encourages the UAV to manage this aerodynamic profile: decelerating to loiter within effective coverage areas to secure sufficient communication rates, while maintaining an economic cruising speed between targets.

\subsubsection{Microscopic Speed Adaptation}
\begin{figure}[!t]
    \centering
    \subfloat[Box plot of per-step speed.]{\includegraphics[width=0.49\columnwidth]{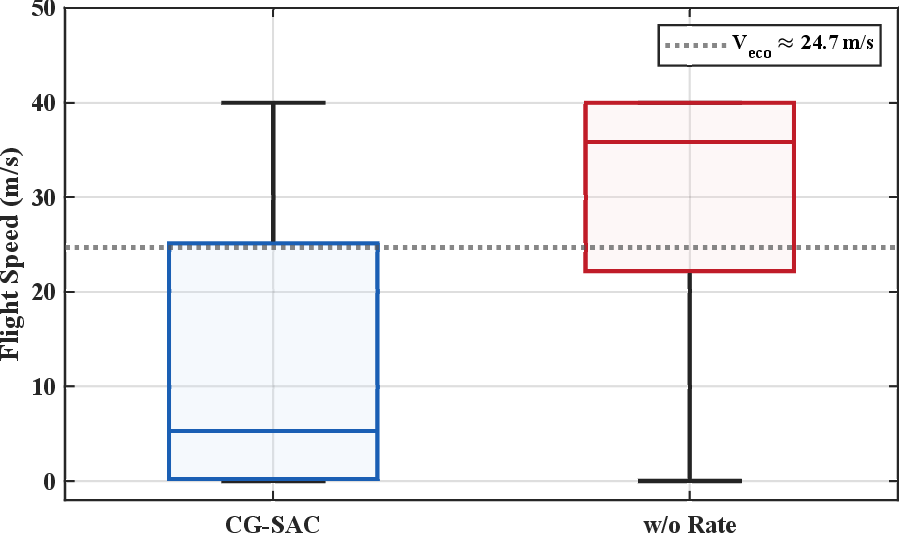}%
    \label{fig:abl_box:boxplot}}
    \hfil
    \subfloat[CDF of per-step speed.]{\includegraphics[width=0.47\columnwidth]{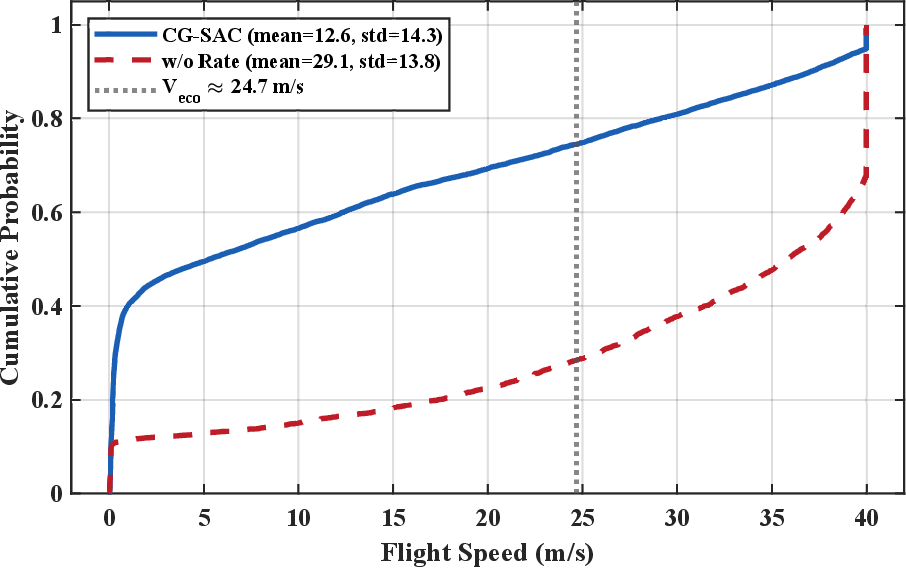}%
    \label{fig:abl_box:cdf}}
    \caption{Microscopic per-step flight-speed distributions of the Full model and the w/o Rate variant over $100$ test episodes. (a) Box plot illustrating the interquartile speed ranges. (b) Cumulative distribution function showing the shift toward high-consumption flight regimes in the ablated variant.}
    \label{fig:abl_box}
\end{figure}

To substantiate this aerodynamic trade-off at a microscopic level, Fig.~\ref{fig:abl_box} contrasts the per-step speed distributions of the Full model and the w/o Rate variant over $100$ test episodes. It is important to note that while the speed statistic in Fig.~\ref{fig:convergence:speed} represents the episode-level average cruising speed, Fig.~\ref{fig:abl_box} captures the full per-step speed distribution, including the low-speed periods required for service interactions. For a single UAV, optimizing service quality necessitates physical deceleration to prolong the interaction time with ground nodes. As shown in the box plot and CDF, the per-step speed distribution has a median of approximately $5$~m/s with a wide interquartile range and an overall per-step mean speed of $12.6$~m/s. This indicates an intelligent, highly dynamic policy: the UAV decelerates when approaching a service point to improve the achievable service quality, while accelerating during transit.

Conversely, without the rate penalty, the median speed of w/o Rate shifts aggressively to $36$~m/s, with its distribution heavily concentrated above the economic speed $V_{\mathrm{eco}}$ and pushing the mean speed up to $29.1$~m/s. A Mann-Whitney U test confirms that this behavioral divergence is statistically significant ($p<0.001$). Consequently, the per-step instantaneous EE of the Full model ($0.822$~Mbits/J) substantially outperforms that of the w/o Rate variant ($0.706$~Mbits/J), demonstrating that explicit QoS constraints are required to suppress energy-wasting sprinting behaviors.

In summary, the ablation study confirms that the selected reward components play complementary roles in mission feasibility and energy efficiency for navigating the highly coupled constraints of a single-UAV ISAC mission. The navigation shaping $R_{\mathrm{nav}}$ provides the primary spatial guidance; the sensing penalty $\mathrm{Penalty}_{\mathrm{sens}}$ promotes stable target-oriented behavior under the sensing requirement; and the rate penalty $\mathrm{Penalty}_{\mathrm{rate}}$ together with the EE reward $\lambda_{\mathrm{EE}}$ encourages energy-efficient flight. Together, these components provide a practical balance among mission completion, service quality, and propulsion efficiency.

\subsection{Sensitivity Analysis of Onboard Energy Budget}
\label{subsec:sensitivity}
To evaluate the sensitivity of the proposed framework to onboard energy limitations, we conduct a sensitivity analysis on the onboard energy budget $E_{\max}$. For this evaluation, $E_{\max}$ is directly enforced as an additional episode-termination condition; an episode terminates with failure once the accumulated energy exceeds the prescribed budget, while the nominal $N_{\max}=800$ step cap remains unchanged. We scale the nominal battery capacity ($212$~kJ) down from $100\%$ to approximately $56\%$ (i.e., $120$~kJ) and evaluate the policies over $100$ unseen scenarios. Error bars represent the $95\%$ confidence intervals.

\begin{figure}[!t]
    \centering
    \subfloat[Mission completion rate.]{\includegraphics[width=0.48\columnwidth]{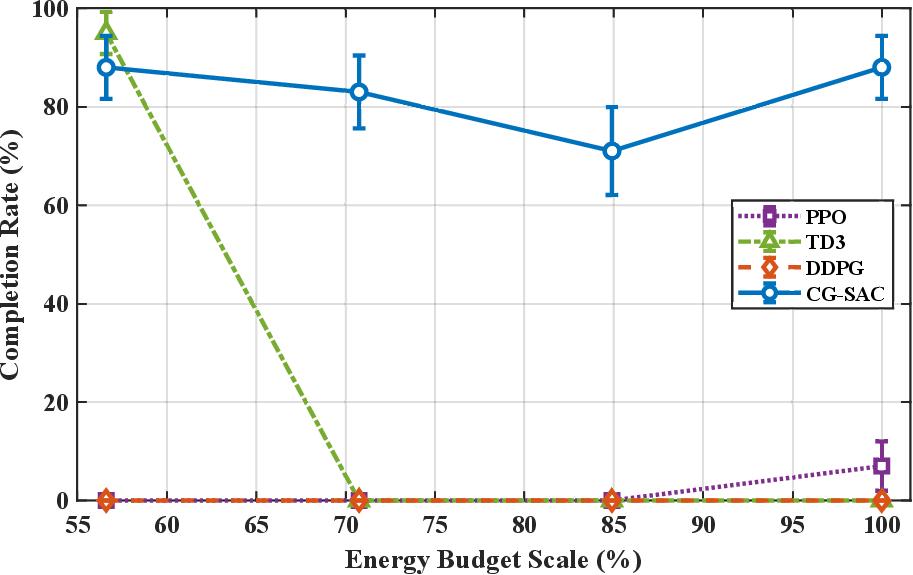}%
    \label{fig:sensitivity:completion}}
    \hfil
    \subfloat[Average energy efficiency.]{\includegraphics[width=0.48\columnwidth]{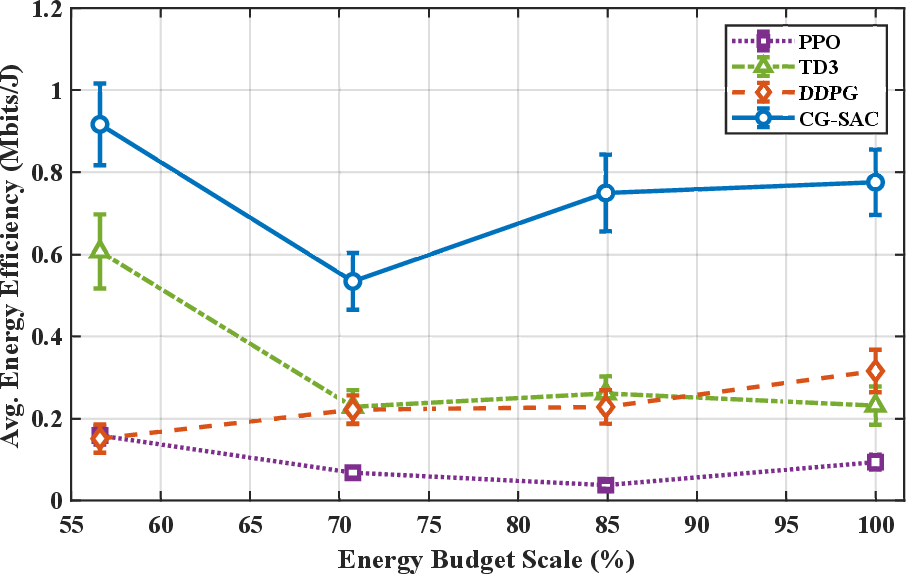}%
    \label{fig:sensitivity:ee}}\\
    \subfloat[Average flight speed.]{\includegraphics[width=0.48\columnwidth]{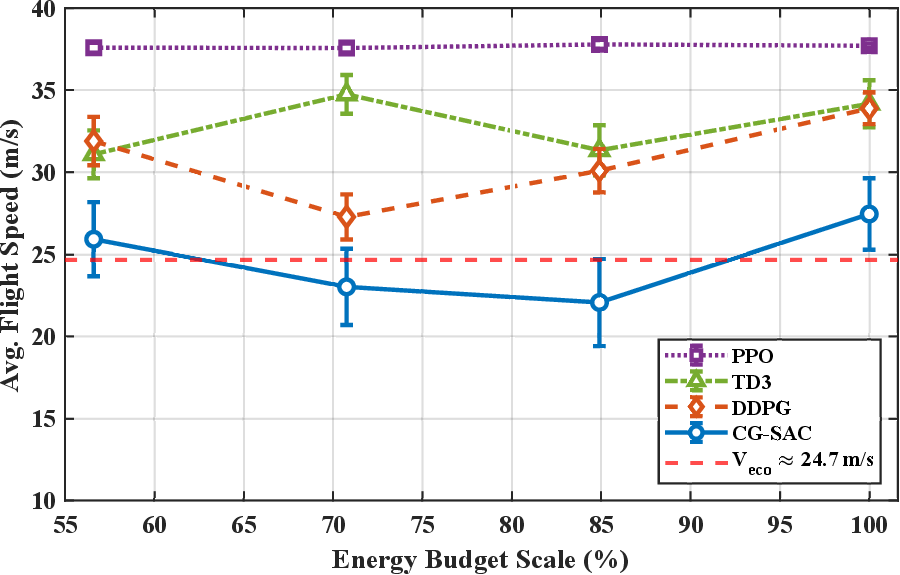}%
    \label{fig:sensitivity:speed}}
    \caption{Sensitivity analysis of the proposed CG-SAC and baseline algorithms under varying onboard energy budgets ($E_{\max}$) with $95\%$ confidence intervals. (a) Mission completion rate. (b) Average energy efficiency. (c) Average flight speed.}
    \label{fig:sensitivity}
\end{figure}

The performance of CG-SAC exhibits a non-monotonic trend over the tested energy-budget range. When the energy budget is abundant ($100\%$), CG-SAC achieves its peak energy efficiency of $0.72$~Mbits/J. As the budget is restricted to $71\%$, the overall EE experiences a dip to roughly $0.53$~Mbits/J. Interestingly, under the most severe constraint of $56\%$, the completion rate returns to nearly $90\%$ and the EE recovers to peak at approximately $0.9$~Mbits/J. The apparent recovery of EE at the most restrictive budget is partly attributable to early termination of energy-infeasible trajectories, which concentrates successful episodes on efficient flight regimes. Throughout this restriction, CG-SAC consistently regulates its average flight speed near the model-derived propulsion-economic speed reference of $24.7$~m/s (Fig.~\ref{fig:sensitivity:speed}), indicating stable adaptation around the economic-speed region.

In contrast, the baseline algorithms exhibit structural instabilities. As shown in Fig.~\ref{fig:sensitivity:speed}, PPO adopts an energy-wasting sprinting strategy near the maximum speed ($37\sim38$~m/s) regardless of the energy budget, resulting in near-zero completion rates and negligible EE. DDPG similarly struggles with excessive speeds and fails to complete the sequential visits. Interestingly, TD3 exhibits an anomalous behavioral shift: while it completely fails at energy budgets of $71\%$ and above due to severe navigational wandering, the extreme energy constraint at the $56\%$ scale appears to inadvertently restrict its wandering, resulting in an unexpected spike in completion rate ($\sim95\%$) and EE. Nevertheless, the proposed curriculum-guided framework maintains the most favorable overall performance across the tested nominal and reduced-energy budgets. Notably, the policy remains effective across these externally imposed energy budgets without explicit remaining-energy feedback in the state. This result indicates that the learned policy is relatively insensitive to the tested budget variations. This behavior is consistent with the combined design of maximum-entropy exploration and the EE-centric composite reward, allowing the agent to dynamically reshape its aerodynamic and trajectory profile under severe hardware limitations.

\subsection{Limitations and Open Directions}
\label{subsec:limitations}

The proposed curriculum-guided reinforcement learning framework establishes a baseline for energy-efficient UAV-ISAC systems by jointly optimizing the 3D trajectory and resource allocation of a single aerial platform. Building upon this foundation, several directions can be extended for future research. 

First, while this study resolves the highly coupled aerodynamic and QoS constraints for a single UAV, it provides a scalable algorithmic structure for multi-UAV cooperative ISAC systems. Future work can extend the current single-agent continuous control to multi-agent frameworks, addressing inter-UAV interference and joint task assignment in large-scale disaster areas~\cite{lu2024tenchallenges,hu2025uavisacsurvey}. 

Second, the designed spatial-temporal navigation policy relies on ideal CSI and static sensing targets to validate the theoretical aerodynamic trade-offs. This deterministic environment modeling serves as a basis for investigating robust policies under imperfect CSI and tracking mobile targets, which are critical for vehicular and low-altitude applications. 

Third, the current model leverages the 3D mobility of the UAV to proactively optimize channel capacity. This geometric optimization can be further synergized with RIS technologies. By jointly optimizing the UAV trajectory and RIS phase shifts, future studies can alleviate the sensing and communication bottleneck in severely obstructed environments. 

Finally, the formulation of the normalized composite reward successfully disentangles the competing ISAC constraints with empirically tuned weights. This modular reward structure naturally paves the way for automated reward shaping or multi-objective reinforcement learning, enabling the agent to adaptively balance communication, sensing, and energy objectives for online deployment.

\section{Conclusion}
\label{sec:conclusion}

This paper investigated energy-efficient mission planning for UAV-ISAC systems through a curriculum-guided soft actor-critic (CG-SAC) framework with propulsion-aware reward shaping, jointly optimizing the 3D trajectory, communication-sensing power split, and per-user power allocation under coupled mobility, communication, sensing, and energy requirements. A speed-dependent rotary-wing propulsion model was used to derive a closed-form horizontal propulsion-economic cruising-speed reference, while a log-linear curriculum progressively tightened the service requirements during training. Extensive evaluations over 2000 randomized scenarios show that CG-SAC achieves an average energy efficiency of 0.72~Mbits/J and requires 107.6 steps on average among successfully completed missions, corresponding to a 66\%--82\% reduction in flight steps relative to the evaluated DRL baselines. At service instants, the learned policy achieves a 99.6\% communication-rate satisfaction ratio while satisfying the adopted sensing-feasibility threshold. The results further reveal mission-aware speed adaptation, with the UAV decelerating near service points and accelerating during transit, highlighting the benefit of jointly accounting for propulsion cost and service requirements in UAV-ISAC mission control.

\bibliographystyle{IEEEtran}
\bibliography{IEEEabrv,reference}

% \begin{IEEEbiography}[{\includegraphics[width=1in,height=1.25in,clip,keepaspectratio]{./pic/TY_Guo.eps}}]{Tai-You Guo}
%     received the B.S. degree in Computer Science and Information Engineering from Ming Chuan University, Taoyuan, Taiwan, in 2024, and the M.S. degree in Communications Engineering from National Chung Cheng University, Chiayi, Taiwan, in 2026. His research interests include deep reinforcement learning, UAV communications, integrated sensing and communication (ISAC), and energy-efficient trajectory and resource optimization.
% \end{IEEEbiography}

% \begin{IEEEbiography}[{\includegraphics[width=1in,height=1.25in,clip,keepaspectratio]{./pic/CC_Lai.eps}}]{Chuan-Chi Lai}
%     (Member, IEEE) received the Ph.D. degree in Computer Science and Information Engineering from the National Taipei University of Technology, Taiwan, in 2017. He held research and faculty positions at National Chiao Tung University and Feng Chia University prior to his current role. Since 2024, he has been an Assistant Professor with the Department of Communications Engineering, National Chung Cheng University, Minxiong Township, Chiayi County, Taiwan. His research interests include mobile edge computing, UAV networks, and AI for wireless communications. Dr. Lai was a recipient of the Postdoctoral Researcher Academic Research Award from the NSTC, Taiwan, in 2019, and Best Paper Awards at WOCC (2018, 2021) and ICUFN (2015).
% \end{IEEEbiography}

\end{document}